\documentclass[manuscript, nonacm]{acmart}
\AtBeginDocument{%
  }

\setcopyright{acmlicensed}
\copyrightyear{2018}
\acmYear{2018}
\acmDOI{XXXXXXX.XXXXXXX}
\acmConference[Conference acronym 'XX]{Make sure to enter the correct
  conference title from your rights confirmation email}{June 03--05,
  2018}{Woodstock, NY}
\acmISBN{978-1-4503-XXXX-X/2018/06}

\usepackage{todonotes}
\usepackage{xltabular}
\usepackage{makecell}
\usepackage{multirow}
\usepackage{booktabs}
\usepackage{xcolor}
\DeclareRobustCommand{\eg}{\textit{e.g.}}
\DeclareRobustCommand{\ie}{\textit{i.e.}}

\begin{document}

\title[The Landscape of Large Language Model-based Affective Personal Informatics]{Scoping the Landscape of Large Language Model-based Affective Personal Informatics: Review, Reflection, Future Direction}

\author{Minjie Liu}
\email{minjie.liu@unsw.edu.au}
\affiliation{%
  \institution{The University of New South Wales}
  \city{Sydney}
  \country{Australia}}

\authornote{Both authors contributed equally to this research.}
\author{Lihuan Li*}
\email{lihuan.li@student.unsw.edu.au}
\author{Xuran Li*}
\email{xuran.li@unsw.edu.au}
\affiliation{%
  \institution{The University of New South Wales}
  \city{Sydney}
  \country{Australia}}

\author{Flora Salim}
\email{flora.salim@unsw.edu.au}
\affiliation{%
  \institution{The University of New South Wales}
  \city{Sydney}
  \country{Australia}}  

\author{Benjamin Tag}
\email{benjamin.tag@unsw.edu.au}
\authornote{Corresponding author.}
\affiliation{%
  \institution{The University of New South Wales}
  \city{Sydney}
  \country{Australia}}

\renewcommand{\shortauthors}{Liu et al.}

\begin{abstract}
Affective personal informatics (PI) involves capturing, tracking, analyzing, and reflecting on personal data to support emotional self-awareness and mental well-being.
Recent advances in large language models (LLMs) are reshaping affective PI, yet a comprehensive survey of LLM-based affective PI systems is lacking.
To bridge this gap, we systematically analyzed 45 papers published between January 2020 and June 2026 across four major computing and health databases, examining their macro trends, system taxonomies, and research methodologies.
Our synthesis reveals an overreliance on multi-stage, text-centric pipelines, a predominance of active, screen-bound interactions, and a severe demographic skew toward young, university-campus-based cohorts in empirical evaluations.
Based on these findings, we critically reflect on prevailing architectural and methodological shortcomings and outline actionable future directions for building end-to-end multimodal, low-burden, and inclusive affective PI systems.
\end{abstract}




\begin{teaserfigure}
\centering
\includegraphics[width=1\textwidth]{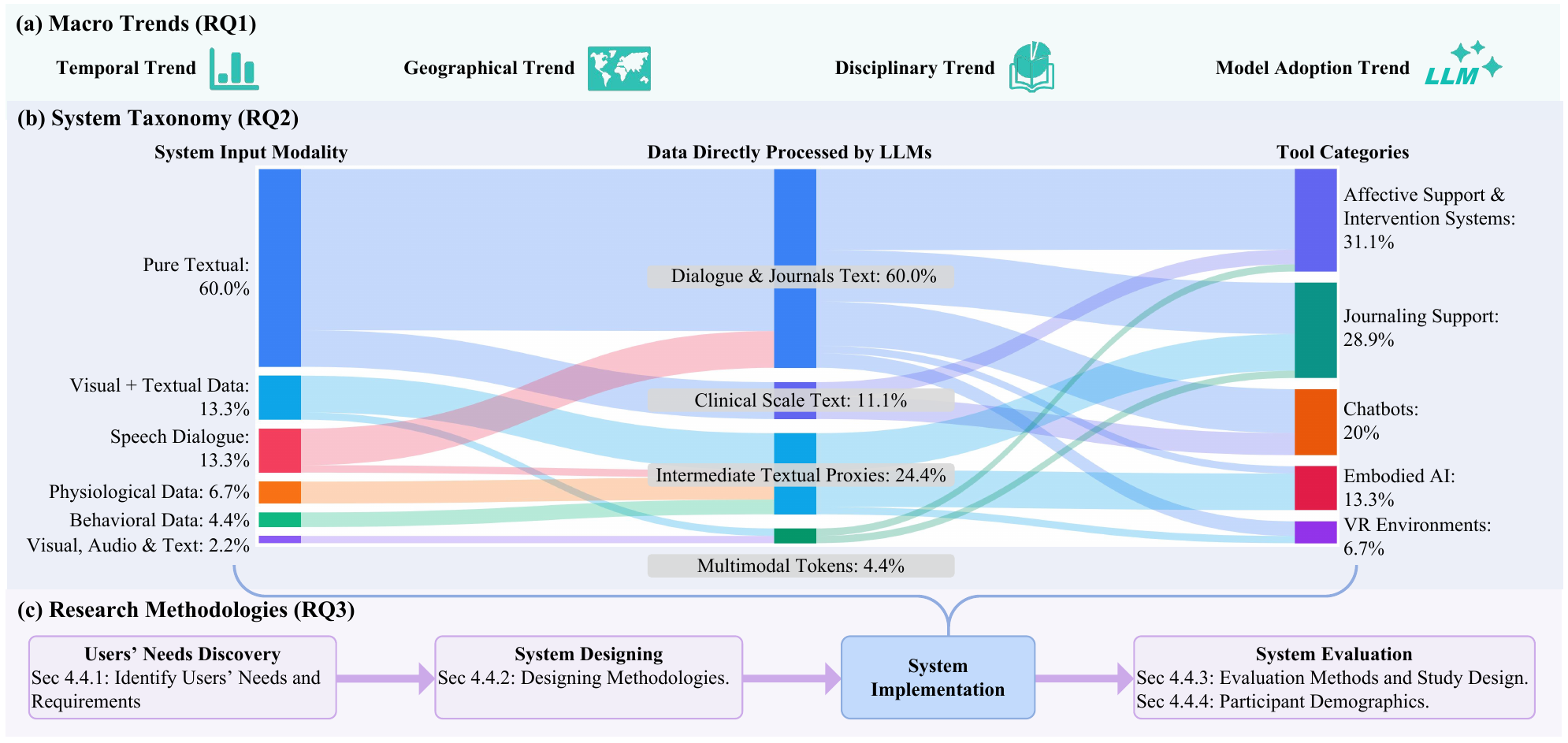}
\caption{Overview of our scoping review across three RQs: \textbf{(a) Macro trends (RQ1)} characterizes temporal, geographical, disciplinary, and model adoption trends; \textbf{(b) System taxonomies (RQ2)} maps the data flow from input modalities and LLM representations to tool categories; and \textbf{(c) Research methodologies (RQ3)} maps reviewed studies onto four stages of the HCI system lifecycle, with system implementation corresponding to the taxonomy in RQ2.}
\label{fig: teaser figure}
\centering
\end{teaserfigure}


\maketitle


\section{Introduction}
Affective personal informatics (PI) has emerged at the intersection of affective computing and personal informatics~\cite{epstein2015lived}, dedicated to helping individuals capture, track, analyze, and reflect on affective states (\eg, emotion, mood, and stress) to foster emotional self-awareness and promote mental well-being. Early work in affective PI functioned primarily as recording tools, capturing physiological signals and self-reported moods, and visualizing these metrics to facilitate self-monitoring and emotional reflection~\cite{staahl2009experiencing, mcduff2012affectaura}. Recently, the paradigm in affective PI has progressively transitioned from isolated tracking toward diverse interaction forms~\cite{zhou2024squeezable}, contextualized emotional reflection~\cite{li2011understanding}, and personalized intervention~\cite{kocielnik2018reflection}. Together, these developments position affective PI not only as a means of tracking affective states, but as a means of supporting reflection and action in everyday emotional self-care.

Recent advances in large language models (LLMs)~\cite{gpt3, gpt-4o, llama3} have significantly influenced research across many different domains, including human-computer interaction (HCI) and affective computing. Affective PI is no exception, with researchers increasingly exploring how LLMs can extend existing systems and enable new forms of interaction and support. 
This is because LLMs can interpret language in context and generate conversational responses, allowing systems to make sense of fragmented, subjective, and unstructured self-reports that are difficult to process with traditional rule-based methods or classifiers~\cite{dechant2025future, yu2026mindfeed}. These capabilities create opportunities for affective PI systems to support more flexible, contextualized, and personalized forms of emotional reflection and support.

Despite growing interest, there is no comprehensive review of existing literature on LLM-based affective PI tools. Synthesizing this emerging literature is important for understanding how LLMs are used in affective PI, identifying structural limitations in current approaches, and informing future research and design. 
Existing surveys at the intersection of AI and mental well-being have primarily focused on emotion-related computational tasks~\cite{shou2025multimodal, zhang2024sentiment}, the capacity of LLMs to demonstrate empathy~\cite{sorin2024large}, or the evaluation of LLMs from clinical and psychiatric diagnostic perspectives~\cite{wu2026affective, guo2024large}. However, none provide a dedicated synthesis of how LLMs are designed and integrated into affective PI tools.

To address this gap, we conducted a scoping review of 45 papers published between January 2020 and June 2026 across four major computing and health databases, examining research on LLM-based affective PI tools. Specifically, our review is guided by three research questions (RQs): 

\textbf{- RQ1 (macro trends):} What are the temporal, geographical, disciplinary, and model adoption trends of LLMs in affective PI? 

\textbf{- RQ2 (system taxonomies):} How are current LLM-based affective PI tools designed in terms of multimodal data input, processing architecture, and interaction approach? 

\textbf{- RQ3 (research methodologies):} What design and empirical research methods are employed across the research lifecycle of LLM-based affective PI systems, from requirements gathering to system evaluation?


Our synthesis of the 45 reviewed articles reveals three recurring limitations across current tools.
First, existing tools rely heavily on multi-stage, text-centric pipelines, where non-textual data (\eg, physiological and behavioral signals) are converted into intermediate textual proxies, risking information loss and error propagation. 
Second, most tools rely on active, screen-bound interactions, such as text-based journaling and multi-turn conversations, which can impose cognitive and interaction burdens on users. 
Third, empirical evaluations are skewed toward young, educated, university-affiliated cohorts, limiting the generalizability of their findings.
Based on these findings, we discuss current limitations and outline actionable future directions for developing end-to-end multimodal, low-burden, and demographically inclusive affective PI tools.

In summary, this work makes three primary contributions: 
\begin{enumerate}
    \item We provide a scoping review of the macro trends, system taxonomies, and research methodologies across 45 articles of LLM-based affective PI tools; 
    \item We identify three recurring limitations in current tools: reliance on multi-stage text-centric pipelines, active and screen-bound interactions, and empirical evaluations skewed toward young, university-affiliated populations. 
    \item We translate these findings into concrete research opportunities by providing actionable design suggestions toward end-to-end multimodal modeling, low-burden interaction, and inclusive design.
\end{enumerate}

\section{Background and Related Work}
\subsection{Affective PI}
At the intersection of affective computing and PI~\cite{epstein2015lived}, affective PI focuses on helping individuals to systematically track, interpret, and reflect upon their inner psychological experiences, such as emotions, stress, and mood, to foster emotional self-awareness and sustained psychological well-being. Early explorations in this domain concentrated primarily on using personal data (\eg, physiological and visual cues) to support retrospective reflection. For example, Affective Diary~\cite{staahl2009experiencing} paired physiological signals with mobile logs and translated them into expressive visual avatars for emotional reflection. Similarly, AffectAura~\cite{mcduff2012affectaura} integrated multimodal behavioral cues and desktop contexts into an interactive timeline for retrospective reflection.

Currently, the evolution of affective PI research exhibits three prominent trends. First, interaction paradigms have become increasingly diverse, expanding beyond traditional mobile applications to include conversational~\cite{morris2018towards}, tangible~\cite{zhou2024squeezable}, and peripheral~\cite{snyder2015moodlight} interactions. Specifically, \citet{morris2018towards} proposed an empathic conversational agent to support mental health reflection; \citet{zhou2024squeezable} designed a squeezable physical artifact for workplace emotion regulation; and MoodLight~\cite{snyder2015moodlight} dynamically adjusted ambient lighting to convey biosensor-derived arousal states. Second, because of the significance of contextual data in affective interpretation~\cite{picard1997affective}, contextualization has emerged as another key trend in affective PI. For instance, \citet{li2011understanding} contextualized affective metrics using daily situational data (\eg, calendar events, locations, and social activities), while MoodScope~\cite{likamwa2013moodscope} inferred users' mood states directly from background smartphone usage contexts. Third, researchers have increasingly prioritized personalization within affective PI systems: \citet{sarker2014assessing} modeled individuals' dynamic receptivity to stress interventions to personalize the timing of delivery, while \citet{kocielnik2018reflection} tailored conversational prompts and dialogue pacing to an individual's unique interaction habits.

In terms of development, affective PI has utilized a variety of technologies, such as mobile system development~\cite{staahl2009experiencing}, passive sensing~\cite{wang2014studentlife}, virtual reality (VR)~\cite{van2016deep}, classical machine learning (ML) algorithms~\cite{sano2013stress}, and rule-based dialogue systems~\cite{fitzpatrick2017delivering}. However, these conventional technologies present critical limitations: classical classifiers produce coarse, uninterpretable labels, while rule-based dialogue systems are constrained by predefined interaction rules and lack the flexibility to handle nuanced human feelings. In contrast, the emergence of LLMs~\cite{gpt3, gpt-4o} marks a significant shift. LLMs offer new capabilities by interpreting and integrating different forms of emotional data and generating personalized reflections and responses~\cite{zhang2024autojournaling, yang2025humanai}.
These capabilities have led researchers to increasingly explore LLMs in affective PI systems.

\subsection{LLMs in Affective Computing}
Affective computing aims to develop digital systems capable of recognizing, processing, understanding, interpreting, and simulating human affect~\cite{picard1997affective}. Leveraging their advanced reasoning and language generation capabilities, LLMs~\cite{gpt3, gpt-4o, llama3} have been deployed across a diverse range of affective computing domains.
The first category is \textit{affective algorithms}, such as emotion recognition~\cite{dutta2025llm, li2026llms}, emotion reasoning~\cite{cheng2024emotion, lian2025affectgpt}, and emotion-cause pair extraction~\cite{wu2024enhancing}. 
The second category focuses on \textit{computational empathy}~\cite{cuff2016empathy}.
A growing body of research suggests that LLMs can, in specific scenarios, match or even surpass human levels of perceived empathy. For instance, \citet{wang2023emotional} developed a text-based emotional understanding benchmark and found that GPT-4~\cite{gpt4} outperformed 89\% of human participants. Similarly, across two empirical studies, \citet{lee2024large} demonstrated that LLMs could generate peer-support responses perceived as more empathic than those written by humans. Conversely, other work highlights critical limitations, cautioning that this apparent empathy may be superficial; for example, \citet{cuadra2024illusion} showed that LLMs exhibit a significantly poorer grasp of nuanced user experiences than human respondents. Overall, while LLMs display substantial empathetic competence in controlled textual settings, their actual depth of empathy varies considerably across task demands.
\textit{Clinical applications} are the third category of LLM applications in affective computing. For example, \citet{chen2023llm} evaluated LLMs in psychiatric interview simulations, showing their ability to assess multi-dimensional clinical symptom profiles. In a different project, \citet{yang2023towards} presented MentalLLaMA, and employed instruction-tuned LLMs to perform standardized mental health screening while generating clinical interpretations for their assessments. Similarly, \citet{englhardt2024classification} used LLMs to derive clinical insights from multimodal behavioral data.

Distinct from work on \textit{affective algorithms}, the \textit{empathy of conversational agents}, or \textit{clinical diagnostics}, \textit{affective PI} shifts the focus toward supporting everyday, longitudinal emotional well-being for individuals~\cite{zhang2024autojournaling, yang2025humanai}. Instead of formal clinical evaluation~\cite{chen2023llm} or emotion recognition~\cite{dutta2025llm}, LLM-based affective PI focuses on unobtrusively supporting ongoing self-awareness~\cite{yu2026mindfeed}, daily reflective sensemaking~\cite{burns2025remigraph}, and contextual emotional regulation~\cite{li2025healbubble}. Despite rapid progress, existing literature has yet to provide a focused synthesis of how LLMs are used in affective PI tools. We address this gap in this paper through a scoping review of LLM-based affective PI.

\begin{figure*}[t!]
\centering
\includegraphics[width=1\textwidth]{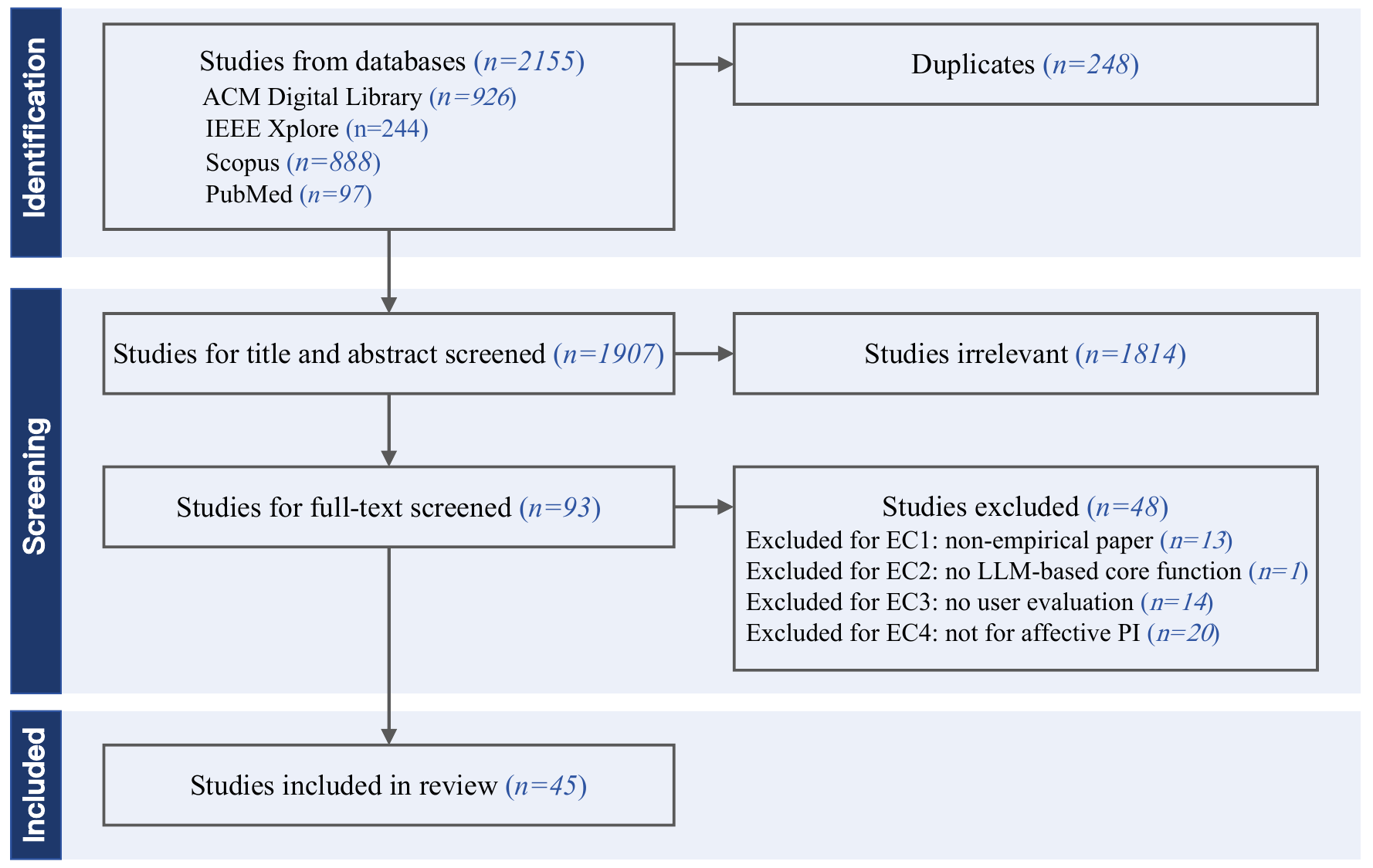}
\vspace{-15pt}
\caption{PRISMA results. In the identification stage, we found 2155 papers from four databases. After removing 248 duplicates, we got 1907 papers for screening. We first did title/abstract screening and concluded with 93 papers remaining. After a full-text assessment, 45 papers constructed the final corpus.}
\vspace{-11pt}
\label{fig: prisma}
\centering
\end{figure*}

\section{Methodology}
We constructed the corpus by following the Preferred Reporting Items for Systematic Reviews and Meta Analysis (PRISMA) guidelines~\cite{page2021prisma}, as shown in Fig.~\ref{fig: prisma}. We utilized the software Covidence\footnote{\url{https://www.covidence.org/}} to manage the review papers.

\subsection{Identification}
To comprehensively search the literature on LLM-based tools for affective PI, we conducted a literature search across three major computer science (CS) databases (ACM Digital Library, IEEE Xplore, and Scopus) and one major healthcare database (PubMed). We designed the search string by combining LLM-related keywords (\eg, Large Language Model and LLM) and affective PI-related keywords (\eg, emotion regulation~\cite{wadley2020digital}, emotion reflection~\cite{li2010stage}, and emotion self-awareness~\cite{calvo2014positive}). Our search timeframe was from January 2020 to June 2026. We set January 2020 as the start date because LLMs emerged following the GPT-2 breakthrough in 2019~\cite{gpt2}. Our preliminary search confirmed that no prior work applied LLMs to affective PI before 2020. We concluded the search at the end of the first half of 2026 (June 2026) to capture a complete semi-annual trend.

We searched the following search string in the title, keywords, abstracts, and full text of papers from January 2020 to June 2026, in each database:

\textit{("LLM" \textbf{OR} "LLMs" \textbf{OR} "Large Language Model" \textbf{OR} "Large Language Models" \textbf{OR} "GPT") \textbf{AND} ("affective personal informatics" \textbf{OR} "emotion regulation" \textbf{OR} "emotion reflection" \textbf{OR} "self-reflection" \textbf{OR} "emotion awareness" \textbf{OR} "emotional awareness" \textbf{OR} "emotion self-awareness" \textbf{OR} "emotional self-awareness" \textbf{OR} "emotion diary" \textbf{OR} "emotional diary" \textbf{OR} "emotion analysis" \textbf{OR} "emotional analysis" \textbf{OR} "stress regulation" \textbf{OR} "stress awareness" \textbf{OR} "stress tracking" \textbf{OR} "mood tracking" \textbf{OR} "mood diary" \textbf{OR} "introspection" \textbf{OR} "metacognition" OR "thought record")}.

We found 926, 244, 888, and 97 articles from the ACM Digital Library, IEEE Xplore, Scopus, and PubMed, respectively, for a total of 2155 articles. After removing 248 duplicates, 1907 papers remained for screening, as of August 31st, 2026.

\subsection{Screening and Eligibility}
To ensure methodological rigor and minimize selection bias, three independent reviewers conducted the screening process using the Covidence platform across two screening phases. In the initial title and abstract screening phase, two reviewers independently screened all identified records against the predefined eligibility criteria. Any conflicting decisions between the two reviewers were resolved through discussions arbitrated by a third reviewer, the first author. To prevent the inadvertent omission of relevant literature, the first author also conducted a secondary audit of all articles excluded during this initial round. Subsequently, all candidate papers advancing to the second phase underwent full-text retrieval and independent evaluation by all three reviewers. Disagreements regarding final study eligibility were resolved through collective discussions until a unanimous consensus was reached.

Prior to screening, all reviewers collaboratively discussed and finalized a strict set of inclusion and exclusion criteria.
If papers met \textbf{ALL} of the following inclusion criteria (IC), they were included:
\begin{enumerate}
    \item \textbf{IC1}: Empirical papers that proposed and developed new interactive tools or evaluated existing interactive tools.
    \item \textbf{IC2}: LLMs worked as the core component of the tool's functionality.
    \item \textbf{IC3}: Conducted empirical user studies to evaluate the interactive tools.
    \item \textbf{IC4}: The interactive tools were designed for affective PI, which collected users' personal information to help track, analyze, reflect, or regulate personal affective states, \eg, emotion, mood, and stress.
\end{enumerate}
Conversely, if papers met \textbf{ANY} of the following exclusion criteria (EC), they were excluded.
\begin{enumerate}
    \item \textbf{EC1}: Non-empirical papers (\eg, research proposals, prospective papers, or literature reviews) that neither proposed nor developed new interactive tools nor evaluated the existing interactive tools.
    \item \textbf{EC2}: Utilized LLMs only for non-core functions of the interaction tools, \eg, data generation or method evaluation.
    \item \textbf{EC3}: Did not conduct empirical user studies to evaluate the interactive tools.
    \item \textbf{EC4}: The interactive tools were not for affective PI.
\end{enumerate}

As illustrated in Fig.~\ref{fig: prisma}, we started by conducting a title/abstract screening on the 1907 papers and retained 93 papers for full-text assessment. During the full-text assessing stage, we excluded 48 papers in total (13 for EC1, 1 for EC2, 14 for EC3, and 20 for EC4). As a result, the final corpus contained 45 papers.

\subsection{Data Analysis}
To systematise the data extraction and analysis process for reviewing the selected literature, we developed a comprehensive extraction protocol. Our protocol captured three key dimensions across each selected paper to answer our three RQs: (1) macro trend, (2) system taxonomies, and (3) research methodologies. The specific contents in the three dimensions were as follows: 
\begin{itemize}
    \item \textbf{Macro trends:} publication year, the utilised LLM, geographical distribution, publication venue, and discipline domain.
    \item \textbf{System taxonomies:} system input modalities, data representations directly processed by LLMs, the tool categories, core functionalities, and the interaction paradigms.
    \item \textbf{Research methodologies:} need-finding and requirement elicitation methods, designing methodologies, evaluating methods, study designs, and the participant demographics (\ie, sample size, gender ratio, age distribution, and mental health state).
\end{itemize}
To ensure consistency and reliability, we first independently extracted data from each included paper, followed by detailed reviews and discussions to resolve any discrepancies and achieve a robust consensus. We then adopted an explanatory and synthesis-driven approach to analysis: we systematically categorized and cross-compared the extracted variables to map out current technological architectures, theoretical foundations, and evaluation paradigms in LLM-driven affective PI.

\begin{figure*}[t!]
\centering
\includegraphics[width=1\textwidth]{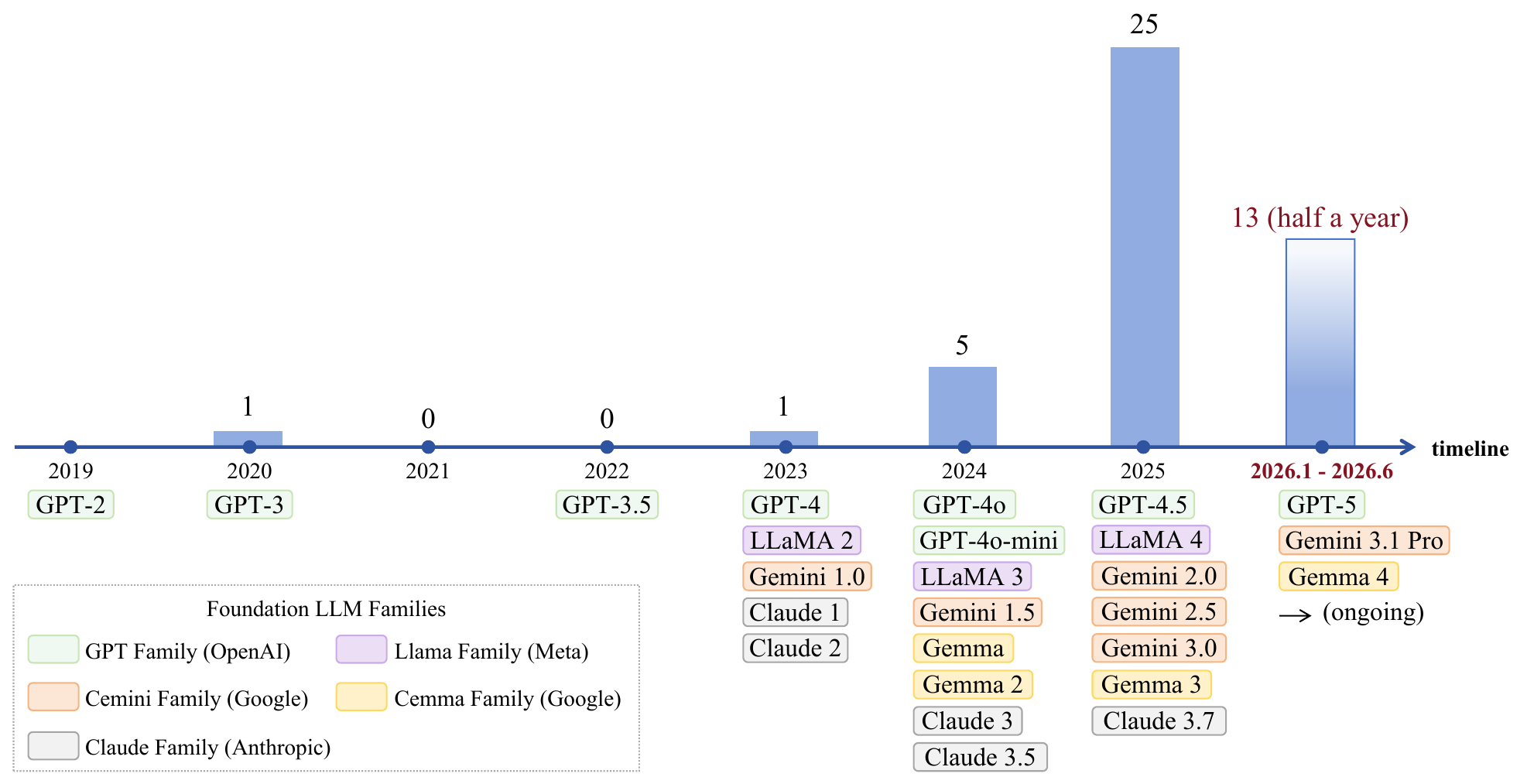}
\caption{Annual distribution of reviewed articles alongside the timeline of foundation LLM releases. The top bar chart illustrates the publication volume per year (January 2020 -June 2026), while the bottom timeline presents major LLM milestones by family. The surge in publications since 2024 reflects the rapid integration of LLMs into affective PI.}
\label{fig: year chart}
\vspace{-20pt}
\centering
\end{figure*}

\section{Results}
\subsection{Temporal Trend and the Adoption of LLMs}
\subsubsection{Temporal Trend in the Context of LLM Advancements}
The publication timeline closely mirrored the technological evolution of LLMs, revealing a clear trajectory from early exploration (2020--2024) to a recent exponential surge (2025--June 2026), as illustrated in Fig.~\ref{fig: year chart}. Before 2022, LLMs, \ie, GPT-2~\cite{gpt2} (2019) and GPT-3~\cite{gpt3} (2020), were still in the early exploratory stage. There was only one paper that applied LLMs to affective PI during 2020-2022, \ie, Mirror Ritual \cite{rajcic2020mirror} (2020). In 2022, the foundation LLM reached its milestone with the release of GPT-3.5. OpenAI's GPT-3.5 greatly promoted the public adoption and the development of LLMs. As shown in the bottom timeline of Fig.~\ref{fig: year chart}, various companies began introducing their own LLM families starting in 2023, \eg, Meta's Llama family, Google's Gemini and Gemma families and Anthropic's Claude family. In 2024, the emergence of multimodal LLMs (MLLMs), \eg, GPT-4o~\cite{gpt-4o}, LLaMA 3~\cite{llama3}, and Gemini 1.5~\cite{gemini1.5}, became another milestone in the development of LLMs. MLLMs enabled fine-grained reasoning over multimodal data, which presented more opportunities for processing complex personal affective data. Compared to the development of foundational LLMs, the application of LLMs in the research of affective PI had lagged behind. There existed one reviewed paper published in 2023~\cite{ni2023focusing} and five papers published in 2024~\cite{nepal2024mindscape, qin2024aiheal, liu2024moodmonster, zhang2024autojournaling, jain2024role}. The research in LLMs' application in affective PI culminated in a significant boom in 2025 ($n = 25$), which accounted for over half (55.6\%) of the entire reviewed corpus. The upward trajectory remained remarkably robust into 2026, with 13 papers recorded in just the first half of the year (January--June 2026), 2026.
This sharp upward trend stressed how continuous breakthroughs in foundation LLM architectures dynamically drove and accelerated methodological innovation across affective computing and personal informatics, reinforcing the need for a review such as ours to map this emerging field.

\begin{table*}[t!]
\centering
\small
\setlength{\tabcolsep}{6pt}
\resizebox{\textwidth}{!}{%
\begin{tabular}{>{\raggedright\arraybackslash}p{0.18\linewidth} >{\raggedright\arraybackslash}p{0.13\linewidth} >{\raggedright\arraybackslash}p{0.28\linewidth} >{\raggedright\arraybackslash}p{0.35\linewidth}}
\toprule
\textbf{LLM Major Family} & \textbf{\# of Studies} & \textbf{Specific Model} & \textbf{References} \\
\midrule
\multirow{7}{*}{\makecell[l]{\textbf{GPT Family}}}
& \multirow{7}{*}{\makecell[c]{29 (64.4\%) \\Duplicates\\ were\\ removed.}} & GPT-2 ($n=2$) & \cite{ni2023focusing, rajcic2020mirror} \\
\cmidrule{3-4}
& & GPT-3 ($n=2$)& \colorbox{gray!20}{\cite{kumar2025large, maples2026how}}\\
\cmidrule{3-4}
& & GPT-3.5 ($n=6$)&  \cite{zisquit2025ai, dechant2025future, laban2026robot, nguyen2026beyond, zhong2026multimodal}, \colorbox{gray!20}{\cite{zhang2025exploring}}, \\
\cmidrule{3-4}
& & GPT-4 ($n=8$)& \cite{nepal2024mindscape, yu2026mindfeed, park2025reimagining, kim2025gpt4}, \colorbox{gray!20}{\cite{kumar2025large, liu2024moodmonster, maples2026how, yang2025humanai}}\\
\cmidrule{3-4}
& & GPT-4o / 4o-mini / 4.1-mini ($n=13$)& \cite{held2025ai, perogamvros2025cathartic, zhang2026gpt, zhu2026memorylib, hwang2026augmentiary, park2025how, li2025healbubble, ikegami2026thank, burns2025remigraph, nunez2025searching}, \colorbox{gray!20}{\cite{liu2024moodmonster, zhang2025exploring, yang2025humanai}}\\
\cmidrule{3-4}
& & GPT-5 ($n=1$) & \cite{kim2026using} \\
\cmidrule{3-4}
& & Unspecified GPT ($n=2$)& \cite{zhao2025research, qin2024aiheal} \\
\midrule
\multirow{2}{*}{\makecell[l]{\textbf{Llama Family}}}
& \multirow{2}{*}{4 (8.9\%)} & LLaMA 2 ($n=1$) & \cite{premalatha2025personalized} \\
\cmidrule{3-4}
& & LLaMA 3 ($n=3$) & \cite{laban2026robot, vm2026serenitymind, srivastava2026autismcarebot} \\
\midrule
\multirow{2}{*}{\makecell[l]{\textbf{Gemini Family}}} 
& \multirow{2}{*}{3 (6.7\%)} & Gemini 1.5 ($n=2$)& \cite{phairoh2025craftjai, zhang2024autojournaling} \\
\cmidrule{3-4}
& & Gemini 2.0 ($n=1$) & \cite{ps2025neurosphere} \\
\midrule
\multirow{4}{*}{\makecell[l]{\textbf{Other Families}}} 
& \multirow{4}{*}{4 (8.9\%))} & Claude 3.5 ($n=1$) & \cite{tsuyuguchi2026futureletter} \\
\cmidrule{3-4}
& & Gemma 3 ($n=1$) & \cite{mora2026refugio} \\
\cmidrule{3-4}
& & Baidu Qianfan ($n=1$) & \cite{he2026mirror} \\
\cmidrule{3-4}
& & GLM-4 ($n=1$) & \cite{liu2025theraissist} \\
\midrule
\textbf{Not Mentioned} 
& 5 (11.1\%) & / & \cite{ahmed_design_2025, wang2025designing, liu2026ragesense, juchli2026digital, jain2024role} \\
\bottomrule
\end{tabular}%
}
\caption{Distribution of Large Language Models utilized across the reviewed articles. Shaded citations \colorbox{gray!20}{[ ]} indicate tools that use more than one LLM.}
\vspace{-20pt}
\label{tab:llm_model_distribution}
\end{table*}

\subsubsection{Distribution of Utilized LLMs}
Table~\ref{tab:llm_model_distribution} details the foundational LLMs powering the reviewed affective PI systems, revealing an ecosystem heavily dominated by OpenAI's GPT family while gradually diversifying towards open-source and multimodal alternatives. Specifically, the GPT family served as the primary backbone in $64.4\%$ of the corpus ($n = 29$, deduplicated), exhibiting an evolutionary shift from early exploratory implementations using GPT-2 ($n = 2$) and GPT-3 ($n = 2$), through conversational baselines powered by GPT-3.5 ($n = 6$) and GPT-4 ($n = 8$), to the current state of the art dominated by the multimodal, low-latency GPT-4o series ($n = 13$), along with frontier explorations of GPT-5 ($n = 1$) and unspecified GPT variants ($n = 2$). Concurrently, researchers increasingly integrated open-source and competing proprietary models to accommodate data privacy, local robotics deployment, and long-context processing; these included Meta's LLaMA family ($8.9\%$, $n = 4$; LLaMA 2: $n = 1$; LLaMA 3: $n = 3$), Google's Gemini family ($6.7\%$, $n = 3$; Gemini 1.5: $n = 2$; Gemini 2.0: $n = 1$), and other alternatives ($8.9\%$, $n = 4$; Claude 3.5: $n = 1$; Gemma 3: $n = 1$; Baidu Qianfan: $n = 1$; GLM-4: $n = 1$), while five studies ($11.1\%$) omitted specific model details. Crucially, a growing macro trend was the adoption of multi-LLM architectures and cross-model comparative benchmarks ($11.1\%$, $n = 5$, shaded in Tab.~\ref{tab:llm_model_distribution}), where authors orchestrated multi-agent pipelines or evaluated iterative generation disparities (\eg, comparing GPT-3.5/4 with GPT-4o or contrasting GPT against open-source LLaMA) to systematically enhance empathetic alignment and therapeutic reliability~\cite{kumar2025large, maples2026how}.

\begin{table*}[t!]
\centering
\small
\setlength{\tabcolsep}{6pt}
\resizebox{\textwidth}{!}{%
\begin{tabular}{>{\raggedright\arraybackslash}p{0.15\linewidth} >{\raggedright\arraybackslash}p{0.12\linewidth}>{\raggedright\arraybackslash}p{0.15\linewidth} >{\raggedright\arraybackslash}p{0.1\linewidth} >{\raggedright\arraybackslash}p{0.2\linewidth}}
\toprule
\textbf{Regions} & \textbf{\# of Studies} & \textbf{Country} & \textbf{\# of Studies} & \textbf{References} \\
\midrule
\multirow{3}{*}{\textbf{East Asia}}
& \multirow{3}{*}{21 (46.7\%)} 
& China & 11 (24.4\%) & \cite{zhang2026gpt, ni2023focusing, zhao2025research, qin2024aiheal, liu2024moodmonster, zhu2026memorylib, li2025healbubble, liu2026ragesense, zhong2026multimodal, he2026mirror, liu2025theraissist} \\
\cmidrule{3-5}
& & Republic of Korea & 6 (13.3\%) & \cite{hwang2026augmentiary, park2025how, park2025reimagining, kim2026using, kim2025gpt4, yang2025humanai} \\
\cmidrule{3-5}
& & Japan & 4 (8.9\%) & \cite{ahmed2025design, honda2025communication, ikegami2026thank, tsuyuguchi2026futureletter} \\
\midrule
\multirow{5}{*}{\textbf{Europe}} 
& \multirow{5}{*}{8 (17.8\%)} 
& Sweden & 2 (4.4\%) & \cite{zhang2025exploring, nunez2025searching} \\
\cmidrule{3-5}
& & Switzerland & 2 (4.4\%) & \cite{perogamvros2025cathartic, juchli2026digital} \\
\cmidrule{3-5}
& & UK & 2 (4.4\%) & \cite{dechant2025future, laban2026robot} \\
\cmidrule{3-5}
& & Spain & 1 (2.2\%) & \cite{mora2026refugio} \\
\cmidrule{3-5}
& & Netherlands & 1 (2.2\%) & \cite{yu2026mindfeed} \\
\midrule
\multirow{2}{*}{\textbf{North America}}
& \multirow{2}{*}{7 (15.6\%)} 
& USA & 5 (11.1\%) & \cite{nepal2024mindscape, held2025ai, srivastava2026autismcarebot, burns2025remigraph, maples2026how} \\
\cmidrule{3-5}
& & Canada & 2 (4.4\%) & \cite{kumar2025large, wang2025designing} \\
\midrule
\textbf{South Asia} & 4 (8.9\%) & India & 4 (8.9\%) & \cite{premalatha2025personalized, ps2025neurosphere, vm2026serenitymind, jain2024role} \\
\midrule
\multirow{4}{*}{\textbf{Other Regions}} 
& \multirow{4}{*}{5 (11.1\%)} 
& Australia & 2 (4.4\%) & \cite{rajcic2020mirror, zhang2024autojournaling} \\
\cmidrule{3-5}
& & Singapore & 1 (2.2\%) & \cite{nguyen2026beyond} \\
\cmidrule{3-5}
& & Thailand & 1 (2.2\%) & \cite{phairoh2025craftjai} \\
\cmidrule{3-5}
& & Israel & 1 (2.2\%) & \cite{zisquit2025ai} \\
\bottomrule
\end{tabular}%
}
\vspace{2pt}
\caption{Geographical distribution of reviewed articles across regions and countries.}
\vspace{-20pt}
\label{tab:regions}
\end{table*}

\begin{table*}[t!]
\centering
\small
\setlength{\tabcolsep}{6pt}
\resizebox{\textwidth}{!}{%
\begin{tabular}{>{\raggedright\arraybackslash}p{0.15\linewidth} c >{\raggedright\arraybackslash}p{0.3\linewidth} >{\raggedright\arraybackslash}p{0.5\linewidth}}
\toprule
\textbf{Category} & \textbf{\# of Studies} & \textbf{Venues} & \textbf{References} \\
\midrule
\multirow{6}{*}{\textbf{CS \& HCI}} 
& \multirow{6}{*}{37 (82.2\%)} 
& ACM CHI / CHI EA ($n = 8$) & CHI: \cite{liu2026ragesense, park2025reimagining, rajcic2020mirror}; CHI EA: \cite{park2025how, ikegami2026thank, wang2025designing, srivastava2026autismcarebot, kim2026using}\\
\cmidrule{3-4}
& & IJHCI ($n = 2$) & \cite{yang2025humanai, he2026mirror} \\
\cmidrule{3-4}
& & ACM IMWUT ($n = 3$) & \cite{nepal2024mindscape, zhu2026memorylib, liu2025theraissist} \\
\cmidrule{3-4}
& & ACM DIS ($n = 2$) & \cite{hwang2026augmentiary, nunez2025searching} \\
\cmidrule{3-4}
& & AIxHEART ($n = 2$) & \cite{qin2024aiheal, liu2024moodmonster} \\
\cmidrule{3-4}
& & Others ($n = 20$) & IJHCS~\cite{nguyen2026beyond}, ACM HCI~\cite{kumar2025large}, UIST Adjunct~\cite{burns2025remigraph}, ACM MobiCom~\cite{zhang2024autojournaling}, IEEE TAFFC~\cite{honda2025communication}, IEEE GCCE~\cite{ahmed2025design}, IEEE SMC~\cite{ni2023focusing}, IEEE RO-MAN~\cite{laban2026robot}, AI \& SOCIETY~\cite{maples2026how}, AHs~\cite{tsuyuguchi2026futureletter}, Frontiers in VR~\cite{mora2026refugio}, ICHEC~\cite{yu2026mindfeed}, GAIDMA~\cite{li2025healbubble}, CCAC~\cite{zhong2026multimodal}, ICTC~\cite{zhao2025research}, SSITCON~\cite{ps2025neurosphere}, ICCCES~\cite{vm2026serenitymind}, ECTI-CON~\cite{phairoh2025craftjai}, EngiTek~\cite{premalatha2025personalized}, ICEECT~\cite{jain2024role} \\
\midrule
\multirow{2}{*}{\textbf{Health \& Psychology}} 
& \multirow{2}{*}{8 (17.8\%)} 
& JMIR Formative Research ($n = 3$) & \cite{zisquit2025ai, dechant2025future, zhang2026gpt} \\
\cmidrule{3-4}
& & Others ($n = 5$) & JMIR~\cite{zhang2025exploring}, JMIR Mental Health~\cite{held2025ai}, Journal of Sleep Research~\cite{perogamvros2025cathartic}, Frontiers in Psychiatry~\cite{kim2025gpt4}, SpringerBriefs in Psychology~\cite{juchli2026digital} \\
\bottomrule
\end{tabular}%
}
\vspace{1pt}
\caption{Classification of publication categories and specific venues across reviewed articles.}
\vspace{-20pt}
\label{tab:category}
\end{table*}

\begin{table*}[t!]
\centering
\small
\setlength{\tabcolsep}{5pt}
\begin{tabularx}{\textwidth}{>{\raggedright\arraybackslash}p{0.18\textwidth} >{\centering\arraybackslash}p{0.10\textwidth} >{\raggedright\arraybackslash}p{0.28\textwidth} >{\raggedright\arraybackslash}X}
\toprule
\textbf{System Input Modality} & \textbf{\# of Studies} & \textbf{Data Directly Processed by LLM} & \textbf{References \& Tools} \\
\midrule

\multirow{3}{*}{\makecell[l]{\textbf{Pure Textual}\\ \textbf{Data}}}
& \multirow{3}{*}{27 (60.0\%)}
& Interactive Dialogue Text ($n=14$) & \makecell[l]{
    Socrates 2.0~\cite{held2025ai}, Mindy~\cite{kumar2025large}, Replika~\cite{maples2026how},\\ GYARU-CHAT~\cite{ikegami2026thank}, SerenityMind~\cite{vm2026serenitymind}, \\
    Future Me~\cite{dechant2025future}, Allie~\cite{zhang2026gpt},\\ POCKET-MIND~\cite{yang2025humanai}, \citet{nguyen2026beyond},\\
\citet{ni2023focusing}, AutismCarebot~\cite{srivastava2026autismcarebot},\\
SaaS Companion~\cite{premalatha2025personalized},
    NeuroSphere~\cite{ps2025neurosphere},\\ HealBubble~\cite{li2025healbubble}} \\
\cmidrule{3-4}
& & Clinical Scale, Assessment \& Protocol Text ($n=5$) & \makecell[l]{
    MIND~\cite{kim2026using}, \citet{park2025how}, TrueBalance~\cite{zhang2025exploring},\\ Psychodynamic Evaluator~\cite{kim2025gpt4},\\ Cathartic Dream Analysis~\cite{perogamvros2025cathartic}} \\
\cmidrule{3-4}
& & Textual Journals \& Reflective Narratives ($n = 8$) & \makecell[l]
    {MoodMonster~\cite{liu2024moodmonster}, \citet{jain2024role}, \\
    CRAFTJAI~\cite{phairoh2025craftjai}, \citet{wang2025designing}, \\ 
    \citet{nunez2025searching}, Augmentiary~\cite{hwang2026augmentiary}, \\ Reverba~\cite{zhong2026multimodal}, FutureLetter~\cite{tsuyuguchi2026futureletter}} \\
\midrule

\multirow{3}{*}{\makecell[l]{\textbf{Visual +}\\ \textbf{Textual}\\ \textbf{Data}}}
& \multirow{3}{*}{6 (13.3\%)}
& Vision tokens from continuous screenshot stream ($n=1$) & 
    AutoJournaling~\cite{zhang2024autojournaling} \\
\cmidrule{3-4}
& & Preprocessed VLM image captions \& concept entity graphs ($n=3$) & 
    MindFeed~\cite{yu2026mindfeed}, RemiGraph~\cite{burns2025remigraph}, MemoryLib~\cite{zhu2026memorylib} \\
\cmidrule{3-4}
& & Classified facial emotion labels \& intensity metadata ($n = 2$) & 
    Mirror Ritual~\cite{rajcic2020mirror}, Mirror Face~\cite{he2026mirror} \\
\midrule

\multirow{2}{*}{\makecell[l]{\textbf{Speech}\\ \textbf{Dialogue}}}
& \multirow{2}{*}{6 (13.3\%)}
& ASR-transcribed voice dialogue \& self-talk streams ($n = 5$) & \makecell[l]{
\citet{zisquit2025ai}, \citet{honda2025communication},\\
\citet{zhao2025research}, TherAIssist~\cite{liu2025theraissist},\\ ephoria~\cite{juchli2026digital}} \\
\cmidrule{3-4}
& & ASR-transcribed speech parsed into JSON environment schemas ($n=1$) & 
    Refugio~\cite{mora2026refugio} \\
\midrule

\multirow{2}{*}{\makecell[l]{\textbf{Physiological}\\ \textbf{Data}}}
& \multirow{2}{*}{3 (6.7\%)}
& Estimated EEG/HRV emotion labels injected into prompt ($n=2$) & 
    QTrobot~\cite{laban2026robot}, AI-HEAL~\cite{qin2024aiheal} \\
\cmidrule{3-4}
& & Aggregated biometric time-series (HR) \& context text strings ($n=1$) & 
    CLOUDS~\cite{ahmed2025design} \\
\midrule

\textbf{Behavioral Data}
& 2 (4.4\%)
& Aggregated behavioral summary text (GPS clusters, step counts, screen time) ($n=2$)
& MindScape~\cite{nepal2024mindscape}, Personal Data Probe~\cite{park2025reimagining} \\
\midrule

\makecell[l]{\textbf{Visual, Audio}\\ \textbf{\& Textual Data}}
& 1 (2.2\%)
& Acoustic-sensed Valence-Arousal history + Vision-LLM gameplay frames ($n=1$)
& RageSense~\cite{liu2026ragesense} \\
\bottomrule
\end{tabularx}
\caption{Classification of system input data modality vs. data directly processed by LLM of reviewed papers.}
\vspace{-20pt}
\label{tab:data_modalities}
\end{table*}

\subsection{Geographical and Disciplinary Trend}
The geographical analysis indicates a globally distributed yet regionally concentrated research effort across five main areas, as presented in Tab.~\ref{tab:regions}. East Asia emerged by far as the dominant contributor, accounting for nearly half of the corpus (46.7\%, $n = 21$), driven by research from China ($n = 11$), the Republic of Korea ($n = 6$), and Japan ($n = 4$). Europe followed as the second-largest hub (17.8\%, $n = 8$), comprising contributions from Sweden ($n = 2$), Switzerland ($n = 2$), the UK ($n = 2$), Spain ($n = 1$), and the Netherlands ($n = 1$). North America accounted for 15.6\% ($n = 7$), comprising the USA ($n = 5$) and Canada ($n = 2$). Additionally, South Asia contributed 8.9\% ($n = 4$, entirely from India). Other regions collectively accounted for 11.1\% ($n = 5$), including Australia ($n = 2$), Singapore ($n = 1$), Thailand ($n = 1$), and Israel ($n = 1$). This distribution highlighted a strong leadership of the East Asian academia in the integration of LLMs with affective PI.

Regarding disciplinary distribution indicated in Tab.~\ref{tab:category}, the reviewed literature was predominantly concentrated in Computer Science and Human-Computer Interaction (CS \& HCI), which accounted for 82.2\% of the corpus ($n = 37$), while Health and Psychology venues comprised the remaining 17.8\% ($n = 8$). Within CS \& HCI, ACM SIGCHI-sponsored conferences represented the primary publication hub, with ACM CHI and its Extended Abstracts (CHI EA) collectively contributing 8 papers, followed by flagship venues including ACM IMWUT ($n = 3$), IJHCI ($n = 2$), ACM DIS ($n = 2$), and AIxHEART ($n=2$). Notably, the CHI Extended Abstract (EA) track accounted for a substantial proportion ($n=5$), outnumbering full CHI papers ($n=3$). This highlighted the exploratory pace of LLM innovation, with EA facilitating rapid prototyping and early-stage validation over extended evaluation cycles.

\subsection{System Pipelines: Input Data and Tool Forms}

\subsubsection{System Input Modalities and Data Processing}
\label{sec: system input modalities}
Table~\ref{tab:data_modalities} summarizes the taxonomy of the system input modalities and their corresponding LLM-interpretable intermediate representations across the reviewed corpus. Overall, the reviewed systems spanned six input configurations: pure textual data ($60.0\%$, $n = 27$), visual and textual data ($13.3\%$, $n = 6$), speech dialogue ($13.3\%$, $n = 6$), physiological signals ($6.7\%$, $n = 3$), passive behavioral data ($4.4\%$, $n = 2$), and tri-modal visual, audio, and textual streams ($2.2\%$, $n = 1$).

Pure \textit{textual data} represented the predominant sensing modality ($60.0\%$, $n = 27$), which systems ingested directly through three primary functional formats without cross-modal transformation. The majority ($n = 14$) processed interactive dialogue text in real time to sustain empathetic exchanges, Cognitive Behavioral Therapy (CBT)-based belief revisions, and supportive scaffolding (\eg, Socrates 2.0~\cite{held2025ai}, Mindy~\cite{kumar2025large}, and GYARU-CHAT~\cite{ikegami2026thank}). Another branch ($n = 8$) ingested textual journals and free-form reflective narratives to generate cognitive feedback, extract emotional themes, or prompt expressive writing (\eg, Augmentiary~\cite{hwang2026augmentiary}, Reverba~\cite{zhong2026multimodal}, and MoodMonster~\cite{liu2024moodmonster}). The remaining studies ($n = 5$) processed structured clinical scale protocols and assessment text to conduct conversational symptom screening and psychological report formulations (\eg, automated PHQ-9 tracking in MIND~\cite{kim2026using} and psychodynamic evaluation parsing in \citet{kim2025gpt4}).

\textit{Visual and textual data} accounted for $13.3\%$ of the corpus ($n = 6$), where raw visual streams were translated into LLM-digestible semantic formats via varied preprocessing pipelines. Specifically, AutoJournaling~\cite{zhang2024autojournaling} directly fed multimodal LLMs with raw vision tokens extracted from high-frequency continuous smartphone screenshot streams. In contrast, three studies utilized upstream Vision-Language Models (VLMs) to convert life-logging images and personal data traces into intermediate textual captions and concept entity graphs for downstream memory-graphing and reflection (MindFeed~\cite{yu2026mindfeed}, RemiGraph~\cite{burns2025remigraph}, and MemoryLib~\cite{zhu2026memorylib}). Additionally, Mirror Ritual~\cite{rajcic2020mirror} and Mirror Face~\cite{he2026mirror} preprocessed raw camera feeds into discrete facial emotion classification labels and intensity metadata, which were subsequently injected into prompt contexts.

The remaining four categories instantiated \textit{spoken, biometric, behavioral, and complex multimodal pipelines} ($26.7\%$, $n = 12$ in total). Spoken audio dialogue ($13.3\%$, $n = 6$) converted vocal self-talk and spoken conversations into automatic speech recognition (ASR)-transcribed text streams (\eg, \citet{zisquit2025ai}, \citet{honda2025communication}, and ephoria~\cite{juchli2026digital}), with Refugio~\cite{mora2026refugio} further parsing transcribed speech into structured JSON environment schemas to dynamically render VR safe places. Physiological signals ($6.7\%$, $n = 3$) were incorporated either by injecting discrete electroencephalogram (EEG)/heart rate variability (HRV) emotion classification labels into dialogue prompts to steer social robots and healing spaces (\citet{laban2026robot}, AI-HEAL~\cite{qin2024aiheal}) or by aggregating time-series heart rate metrics into context text strings (CLOUDS~\cite{ahmed2025design}). Passive behavioral sensing ($4.4\%$, $n = 2$) aggregated daily mobility, step counts, and screen time into summarized context strings to minimize manual logging burden (MindScape~\cite{nepal2024mindscape}, Personal Data Probe~\cite{park2025reimagining}). Finally, RageSense~\cite{liu2026ragesense} ($2.2\%$, $n = 1$) integrated a tri-modal sensing pipeline, fusing near-ultrasonic acoustic sensing history (valence-arousal trajectory) with gameplay video frames to trigger real-time, context-aware LLM interventions during mobile gaming.

\clearpage 
{\small
\raggedbottom 
\begin{xltabular}{\linewidth}{>{\raggedright\arraybackslash}p{0.13\linewidth} >{\centering\arraybackslash}p{0.09\linewidth} >{\raggedright\arraybackslash}p{0.18\linewidth} >{\raggedright\arraybackslash}X}

\toprule
\textbf{Tools Categories} & \textbf{\# of Studies} & \textbf{Specific Tool Classifications} & \textbf{References \& Tool Overview} \\
\midrule
\endfirsthead

\multicolumn{4}{c}{{\bfseries Table \thetable\ -- Continued from previous page}} \\[4pt]
\toprule
\textbf{Tools Categories} & \textbf{\# of Studies} & \textbf{Specific Tool Classifications} & \textbf{References} \\
\midrule
\endhead

\midrule \multicolumn{4}{c}{{\bfseries Table \thetable\ -- Continued on next page}} \\[4pt]
\endfoot

\bottomrule
\caption{Classification of tool categories and specific tool implementations across reviewed articles. Shaded citations \colorbox{gray!20}{[ ]} indicates studies evaluating existing tools rather than proposing new methods.} \label{tab:tools_categories} \\
\endlastfoot

\multirow{4}{*}{\makecell[l]{\textbf{Affective Support}\\\textbf{\& Intervention}\\\textbf{Systems}}}
& \multirow{4}{*}{14 (31.1\%)}
& Psychological Theory-Grounded Support ($n=5$) & 
    Socrates 2.0~\cite{held2025ai}: Facilitates Socratic dialogue for cognitive reappraisal and belief revision;\newline
    Mindy~\cite{kumar2025large}: Delivers digital mindfulness and behavioral change coaching;\newline
    Ephoria~\cite{juchli2026digital}: Provides solution-focused mental health coaching conversations;\newline
    AutismCarebot~\cite{srivastava2026autismcarebot}: Delivers neurodiversity-affirming, emotion-first support;\newline
    \colorbox{gray!20}{Psychodynamic Evaluator~\cite{kim2025gpt4}}: Generates psychodynamic psychological evaluation reports\\
\cmidrule{3-4}
& & Creative Interventions ($n = 5$) & 
    HealBubble~\cite{li2025healbubble}: Synthesizes dynamic ``emotion bubbles'' via generative AI for trauma-informed self-regulation;\newline
    TherAIssist \cite{liu2025theraissist}: Combines human-AI co-creative art-making and conversational agents to assist art therapy homework and therapist-client collaboration;\newline
    GYARU-CHAT~\cite{ikegami2026thank}: Leverages a Gyaru cultural persona to foster a positive mindset;\newline
    SerenityMind~\cite{vm2026serenitymind}: Integrates yogic cultural mindset and scientific self-care for mood management;\newline
    RageSense~\cite{liu2026ragesense}: Uses near-ultrasonic acoustic sensing to detect gaming rage \& triggers context-aware nudges\\
\cmidrule{3-4}
& & Emotional AI Companions ($n=2$) & 
    \colorbox{gray!20}{Replika~\cite{maples2026how}}: AI companion supporting socio-emotional learning and metacognition;\newline
    SaaS Companion~\cite{premalatha2025personalized}: Offers emotionally intelligent clarity and productivity coaching \\
\cmidrule{3-4}
& & Computational Assessment ($n=2$) & 
    NeuroSphere~\cite{ps2025neurosphere}: Uses reinforcement learning to optimize focus, mood, and stress;\newline
    Cathartic Dream Analysis~\cite{perogamvros2025cathartic}: Classifies and scores emotional dream catharsis content \\
\midrule

\multirow{4}{*}{\makecell[l]{\textbf{Journaling} \\ \textbf{Support}}}
& \multirow{4}{*}{13 (28.9\%)}
& AI-Guided Writing Assistance ($n = 5$) & 
    \citet{jain2024role}: Integrates an LLM-based counseling system into journaling to provide real-time supportive feedback and session summaries;\newline
    \citet{wang2025designing}: Provides reframing suggestions and discussions;\newline
    Augmentiary~\cite{hwang2026augmentiary}: Generates interpretive feedback and links to past entries;\newline
    FutureLetter~\cite{tsuyuguchi2026futureletter}: Offers sentence/question nudges to resolve writing stalls;\newline
    MindFeed~\cite{yu2026mindfeed}: Generates reflective prompts from archived social media content \\
\cmidrule{3-4}
& & Journal Text-to-Image Generation ($n=3$) & 
    MoodMonster~\cite{liu2024moodmonster}: Extracts emotion keywords from text to synthesize monster images;\newline
    CRAFTJAI~\cite{phairoh2025craftjai}: Transforms diary entries into gamified cartoon mood characters;\newline
    Reverba~\cite{zhong2026multimodal}: Guides memory recall via dialogue \& generates visual memory art \\
\cmidrule{3-4}
& & Journal Auto-Generation ($n=3$) & 
    MindScape~\cite{nepal2024mindscape}: Uses passive sensing (location, activity) for  context-aware check-ins;\newline
    AutoJournaling~\cite{zhang2024autojournaling}: Summarizes continuous smartphone screenshots to infer feelings;\newline
    Personal Data Probe~\cite{park2025reimagining}: Translates personal informatics logs into generative images \\
\cmidrule{3-4}
& & Memory-Graphing ($n=2$) & 
    RemiGraph~\cite{burns2025remigraph}: Converts lifelogs into structured memory graphs for reminiscence;\newline
    MemoryLib~\cite{zhu2026memorylib}: Structures multimodal egocentric traces into event graphs \\
\midrule

\multirow{3}{*}{\textbf{Chatbots}}
& \multirow{3}{*}{9 (20.0\%)}
& Affective Reflection and Regulation ($n=4$) & 
    \citet{nguyen2026beyond}: Guides users via reflective dialogue to construct fine-grained emotion concepts;\newline
    TrueBalance~\cite{zhang2025exploring}: Supports body image self-reflection and cognitive restructuring for eating habits;\newline
    \citet{nunez2025searching}: Translates bodily felt senses into Haiku poetry for embodied introspection;\newline
    \citet{ni2023focusing}: Provides fine-tuned emotion regulation dialogue centered on adolescent needs \\
\cmidrule{3-4}
& & Goal-Oriented \& Positive Psychology Interventions ($n=3$) & 
    Future Me~\cite{dechant2025future}: Facilitates prospection and future-oriented goal pursuit through adaptive prompts;\newline
    Allie~\cite{zhang2026gpt}: Delivers structured PERMA positive psychology exercises for parents of ASD children;\newline
    POCKET-MIND~\cite{yang2025humanai}: Employs dual-prompt dialogues for emotional exploration \& SMART goal attainment \\
\cmidrule{3-4}
& & Conversational Assessment \& Screening ($n=2$) & 
    MIND~\cite{kim2026using}: Conducts conversational depression screening via automated PHQ-9 item tracking;\newline
    \citet{park2025how}: Evaluates academic stress via self-disclosing dialogue \& SISCO-AS questionnaire parsing \\
\midrule

\multirow{2}{*}{\textbf{Embodied AI}}
& \multirow{2}{*}{6 (13.3\%)}
& Smart Devices ($n = 4$) & 
    Mirror Ritual~\cite{rajcic2020mirror}: Smart mirror generating personalized poetry based on facial emotion analysis;\newline
    AI-HEAL~\cite{qin2024aiheal}: Smart cube healing space integrating CBT dialogue, session summaries, and EEG feedback;\newline 
    CLOUDS~\cite{ahmed2025design}: Smartwatch app providing narrative mood insights from context and biometric data;\newline
    Mirror Face \cite{he2026mirror}: Smart mirror providing real-time facial emotion recognition, gesture interaction, and LLM-generated encouraging feedback for workplace emotion regulation\\
\cmidrule{3-4}
& & Embodied Robots ($n=2$) & 
    \colorbox{gray!20}{QTrobot~\cite{laban2026robot}}: Robot-led intervention facilitating emotional expression and cognitive reappraisal dialogue;\newline
    \citet{honda2025communication}: Social robot generating empathetic dialogue guided by EEG/HRV emotion estimates; \\
\midrule

\textbf{VR Environments}
& 3 (6.7\%)
& \textbackslash
& 
    Refugio~\cite{mora2026refugio}: Generates voice-driven, personalized ``safe place'' VR environments for emotion regulation;\newline
    \citet{zhao2025research}: Delivers multimodal therapeutic dialogue via an AI virtual human within VR;\newline
    \citet{zisquit2025ai}: Facilitates VR perspective-taking self-talk therapy with AI counselor guidance\\
\end{xltabular}
}

\subsubsection{Tool Categories and Functional Paradigms}
\label{sec: tool categories}
Table~\ref{tab:tools_categories} outlines the taxonomy of systems and functional paradigms across the reviewed corpus, structured into five overarching categories: affective support and intervention systems ($31.1\%$, $n = 14$), journaling support ($28.9\%$, $n = 13$), conversational chatbots ($20.0\%$, $n = 9$), embodied AI ($13.3\%$, $n = 6$), and immersive VR environments ($6.7\%$, $n = 3$).

\textit{Affective support and intervention systems} constituted the largest category ($31.1\%$, $n = 14$). 
Specifically, five articles built their tools based on professional psychological theories, \ie, Socratic cognitive reappraisal in Socrates 2.0~\cite{held2025ai}, mindfulness coaching in Mindy~\cite{kumar2025large}, solution-focused coaching in Ephoria~\cite{juchli2026digital}, neurodiversity-affirming support in AutismCarebot~\cite{srivastava2026autismcarebot}, and psychodynamic psychological evaluation in ~\citet{kim2025gpt4}. 
In addition, creative interventions ($n = 5$) explored culturally situated and multimodal metaphors, such as Gyaru persona roleplay in GYARU-CHAT~\cite{ikegami2026thank}, and generative visual emotion bubbles in HealBubble~\cite{li2025healbubble}. 
The remaining tools provided socio-emotional learning and metacognition via emotional AI companions ($n = 2$, \ie, Replika~\cite{maples2026how} and SaaS Companion~\cite{premalatha2025personalized}), or enabled computational assessment ($n = 2$) through reinforcement learning wellness optimization (NeuroSphere~\cite{ps2025neurosphere}) and LLM-based cathartic dream analysis (~\citet{perogamvros2025cathartic}).

\textit{Journaling support systems} ($28.9\%$, $n = 13$) augmented personal expressive writing across four specialized workflows. AI-guided writing assistance ($n = 5$) integrated an LLM-based counseling system in journaling~\cite{jain2024role}), offering interactive reframing suggestions~\cite{wang2025designing}, facilitating deeper introspection by generating contextual feedback on past entries (Augmentiary~\cite{hwang2026augmentiary}), overcoming writer's stall via nudging prompts (FutureLetter~\cite{tsuyuguchi2026futureletter}), or deriving reflective prompts from social media feeds (MindFeed~\cite{yu2026mindfeed}). Furthermore, journal text-to-image generation ($n = 3$) transformed textual logs into expressive visual art, including synthetic monster characters in MoodMonster~\cite{liu2024moodmonster}, gamified mood avatars in CRAFTJAI~\cite{phairoh2025craftjai}, and multimodal memory scenes in Reverba~\cite{zhong2026multimodal}. Concurrently, journal auto-generation tools ($n = 3$) minimized user logging burden by leveraging passive sensing in MindScape~\cite{nepal2024mindscape}, continuous smartphone screenshot streaming in AutoJournaling~\cite{zhang2024autojournaling}, and personal data log translations in ~\citet{park2025reimagining}, while memory-graphing approaches ($n = 2$) synthesized egocentric traces into structured reminiscence graphs in RemiGraph~\cite{burns2025remigraph} and MemoryLib~\cite{zhu2026memorylib}.

\textit{Conversational chatbots} ($20.0\%$, $n = 9$) delivered interactive dialogic guidance across three functional areas. 
First, affective reflection and regulation ($n = 4$) leveraged conversation to help users deconstruct, articulate, and cognitively reframe complex emotional states, \eg, constructing fine-grained emotion concepts in \citet{nguyen2026beyond}, and facilitating body-image cognitive restructuring in TrueBalance~\cite{zhang2025exploring}.
Second, goal-oriented and positive psychology interventions ($n = 3$) guided users toward constructive future prospection, strengths cultivation, and actionable self-improvement through structured psychological exercises.
Third, conversational assessment and screening ($n = 2$) streamlined standardized clinical intake via interactive PHQ-9 depression tracking~\cite{phq-9} in MIND~\cite{kim2026using} and academic stress parsing via SISCO-AS~\cite{SISCO-AS} questionnaires in ~\citet{park2025how}.

The remaining two categories are \textit{embodied AI} and \textit{VR environments}. Within \textit{embodied AI} ($13.3\%$, $n = 6$), smart devices ($n = 4$) embedded intelligence into everyday hardware, while embodied robots ($n = 2$), such as QTrobot~\cite{laban2026robot} and social robots~\cite{honda2025communication}, facilitate physical emotional reappraisal guided by physiological cues. Complementarily, \textit{VR environments} ($6.7\%$, $n = 3$) leveraged spatial immersion for emotion regulation, such as immersive perspective-taking self-talk therapy~\cite{zisquit2025ai}.

\begin{table*}[t!]
\centering
\small
\vspace{20pt}
\resizebox{\textwidth}{!}{
\begin{tabular}{p{0.22\linewidth}  p{0.11\linewidth} | p{0.24\linewidth} p{0.11\linewidth} p{0.22\linewidth}}
\toprule
\textbf{Main Paradigm} & \textbf{\# of Studies} & \textbf{Specific Methodology} & \textbf{\# of Studies} & \textbf{References} \\
\hline
\textbf{Evaluation on Existing Tools} & 3 (6.7\%) & \textbackslash & \textbackslash & \cite{laban2026robot, maples2026how, kim2025gpt4}\\
\hline
\multirow{8}{=}{\textbf{Empirical Formative Research}} & \multirow{8}{*}{\makecell[l]{19 (42.2\%)\\Duplicates\\ were\\ removed.}} 
& User Interviews & 8 (17.8\%) & \cite{nepal2024mindscape, zhang2025exploring, dechant2025future, zisquit2025ai, yang2025humanai}, \colorbox{gray!20}{\cite{li2025healbubble, liu2024moodmonster, liu2026ragesense}}\\
\cmidrule{3-5}
& & Quantitative Surveys & 4 (8.9\%) & \cite{qin2024aiheal, kumar2025large}, \colorbox{gray!20}{\cite{li2025healbubble, liu2024moodmonster}}\\
\cmidrule{3-5}
& & Autobiographical Design & 3 (6.7\%) & \cite{nunez2025searching, park2025reimagining, wang2025designing} \\
\cmidrule{3-5}
& & Expert Consultations & 3 (6.7\%) & \cite{held2025ai, mora2026refugio}, \colorbox{gray!20}{\cite{liu2025theraissist}} \\
\cmidrule{3-5}
& & Focus Groups & 2 (4.4\%) & \colorbox{gray!20}{\cite{liu2026ragesense, liu2025theraissist}} \\
\cmidrule{3-5}
& & Technology Probes & 1 (2.2\%) & \cite{hwang2026augmentiary} \\
\cmidrule{3-5}
& & Experience Sampling & 1 (2.2\%) & \cite{ni2023focusing} \\
\cmidrule{3-5}
& & Co-Design Workshops & 1 (2.2\%) & \cite{zhu2026memorylib} \\
\hline
\multirow{2}{=}{\textbf{Non-Empirical Formative Research}} & 23 (51.1\%) & Literature \& Problem-Driven Derivation & 23 (51.1\%) & \cite{rajcic2020mirror, honda2025communication, phairoh2025craftjai, zhao2025research, ps2025neurosphere, vm2026serenitymind, yu2026mindfeed, burns2025remigraph, nguyen2026beyond, zhong2026multimodal, kim2026using, zhang2024autojournaling, ahmed2025design, perogamvros2025cathartic, premalatha2025personalized, park2025how, juchli2026digital, ikegami2026thank, zhang2026gpt, tsuyuguchi2026futureletter, srivastava2026autismcarebot, jain2024role, he2026mirror}\\
\bottomrule
\end{tabular}}
\vspace{1pt}
\caption{Classification of user needs and requirements identification methods of articles. Shaded citations \colorbox{gray!20}{[ ]} indicate studies that adopt multi-method approaches to identify user needs.}
\vspace{-20pt}
\label{tab:needs_identification}
\end{table*}

\subsection{Research Methodologies}
\subsubsection{Identifying Users’ Needs and Requirements}
Regarding the approach to user need identification prior to system design, the reviewed studies fall into three distinct paradigms: evaluation of existing tools ($6.7\%$, $n = 3$)~\cite{laban2026robot, maples2026how, kim2025gpt4}, empirical formative research ($42.2\%$, $n = 19$), and non-empirical derivation ($51.1\%$, $n = 23$).

Overall, $42.2\%$ of the studies ($n = 19$, deduplicated) conducted \textit{empirical formative research}. Among these empirical methodologies, user interviews were the most prevalent ($17.8\%$, $n = 8$). For example, MindScape~\cite{nepal2024mindscape} conducted interviews with five undergraduate students to uncover their journaling habits and needs. Following interviews, quantitative surveys were the second most common approach ($8.9\%$, $n = 4$), utilized to characterize user preferences and baseline symptoms at scale. Additionally, three articles employed autobiographical design ($6.7\%$, $n = 3$), in which authors leveraged first-person lived inquiries. Other formative research methods included expert consultations with clinical practitioners ($6.7\%$, $n = 3$), focus groups ($4.4\%$, $n = 2$), as well as technology probes, experience sampling, and co-design workshops ($2.2\%$, $n = 1$ each).
Notably, four studies~\cite{li2025healbubble, liu2024moodmonster, liu2026ragesense, liu2025theraissist} adopted multi-method empirical triangulations to comprehensively scope user requirements. Specifically, HealBubble~\cite{li2025healbubble} and MoodMonster~\cite{liu2024moodmonster} conducted two-phase formative studies combining online surveys and interviews to uncover user needs; RageSense~\cite{liu2026ragesense} engaged 24 mobile gamers through a combination of focus group discussions and semi-structured interviews to capture preferred real-time intervention strategies; and TherAIssist~\cite{liu2025theraissist} conducted both one-to-one interviews and focus groups with five professional art therapists to understand the design requirements for art therapy homework.

Conversely, $51.1\%$ of the corpus ($n = 23$) relied on \textit{non-empirical formative research}, in which system requirements were derived directly from prior literature, psychological theories, or problem-driven theoretical rationales without pre-deployment user studies.

\begin{table*}[t!]
\centering
\small
\resizebox{\textwidth}{!}{
\begin{tabular}{>{\raggedright\arraybackslash}p{0.35\linewidth} >{\raggedright\arraybackslash}p{0.18\linewidth} >{\raggedright\arraybackslash}p{0.41\linewidth}}
\toprule
\textbf{Designing Method Paradigm} & \textbf{\# of Studies} & \textbf{References} \\
\midrule
Evaluation-Only Studies & 3 (6.7\%) & \cite{laban2026robot, maples2026how, kim2025gpt4} \\
\midrule
Expert \& Developer-Led Design & 31 (68.9\%) & \cite{held2025ai, perogamvros2025cathartic, zhang2026gpt, ahmed2025design, premalatha2025personalized, honda2025communication, phairoh2025craftjai, zhao2025research, liu2024moodmonster, ps2025neurosphere, vm2026serenitymind, yu2026mindfeed, park2025how, li2025healbubble, ikegami2026thank, tsuyuguchi2026futureletter, burns2025remigraph, zhong2026multimodal, juchli2026digital, mora2026refugio, kim2026using, rajcic2020mirror, yang2025humanai, zhang2024autojournaling, dechant2025future, nguyen2026beyond, zhang2025exploring, zisquit2025ai, srivastava2026autismcarebot, jain2024role, he2026mirror} \\
\midrule
User-Centred Design & 7 (15.6\%) & \cite{ni2023focusing, qin2024aiheal, kumar2025large, nepal2024mindscape, hwang2026augmentiary, liu2026ragesense, liu2025theraissist} \\
\midrule
Autobiographical Design & 3 (6.7\%) & \cite{wang2025designing, nunez2025searching, park2025reimagining} \\
\midrule
Co-Design & 1 (2.2\%) & \cite{zhu2026memorylib} \\
\bottomrule
\end{tabular}}
\vspace{1pt}
\caption{Classification of tool designing methods of reviewed articles.}
\vspace{-20pt}
\label{tab:design_methods}
\end{table*}

\subsubsection{Designing Methodologies}
As summarized in Tab.~\ref{tab:design_methods}, excluding three studies ($6.7\%$, $n = 3$)~\cite{laban2026robot, maples2026how, kim2025gpt4} that evaluated existing tools without proposing new interactive systems, the design methodologies across the remaining literature fell into four categories: expert and developer-led design, user-centered design, autobiographical design, and co-design.

The vast majority of the corpus ($68.9\%$, $n = 31$) adopted an expert and developer-led design approach. In this paradigm, systems were primarily designed directly by researchers and domain experts (\eg, clinical psychologists). For example, Socrates 2.0~\cite{held2025ai} was constructed with clinical psychologists to ensure correct therapeutic Socratic dialogue for cognitive reappraisal, while Allie~\cite{zhang2026gpt} was structured by mental health specialists to deliver evidence-based positive psychology interventions for parents of children with autism.
In contrast, $15.6\%$ of the reviewed studies ($n = 7$) explicitly embraced a user-centered design approach. In these works, researchers iteratively shaped system design through formative empirical inquiries and direct user feedback. For instance, MindScape~\cite{nepal2024mindscape} leveraged formative user investigations to iteratively refine its system's prompt-generation logic. Beyond expert- and user-centered design, a smaller subset of studies leveraged autobiographical design and co-design methods. Specifically, three articles ($6.7\%$, $n = 3$) adopted autobiographical design~\cite{wang2025designing, nunez2025searching, park2025reimagining}, in which researchers served as primary users who iteratively probed their self-designed prototypes over extended timeframes. In addition, MemoryLib~\cite{zhu2026memorylib} uniquely employed co-design ($2.2\%$, $n = 1$) by hosting collaborative workshops with target stakeholders to structure multimodal egocentric memory graphs for reflective recall.

\begin{table}[t!]
\centering
\small
\setlength{\tabcolsep}{8pt}
\resizebox{\textwidth}{!}{
\begin{tabular}{>{\raggedright\arraybackslash}p{0.1\linewidth} >{\raggedright\arraybackslash}p{0.3\linewidth} >{\raggedright\arraybackslash}p{0.1\linewidth} >{\raggedright\arraybackslash}p{0.45\linewidth}}
\toprule
\textbf{Evaluation Dimension} & \textbf{Category / Design Type} & \textbf{\# of Studies} & \textbf{References}\\
\midrule
\multirow{3}{*}{\makecell[l]{\textbf{Method}\\\textbf{Paradigm}}} 
& Mixed Methods & 31 (68.9\%) & \cite{nepal2024mindscape, kumar2025large, zhang2025exploring, zisquit2025ai, held2025ai, dechant2025future, zhang2026gpt, honda2025communication, zhao2025research, qin2024aiheal, liu2024moodmonster, ps2025neurosphere, vm2026serenitymind, zhu2026memorylib, yu2026mindfeed, park2025how, li2025healbubble, ikegami2026thank, srivastava2026autismcarebot, tsuyuguchi2026futureletter, burns2025remigraph, nguyen2026beyond, liu2026ragesense, zhong2026multimodal, park2025reimagining, maples2026how, kim2026using, rajcic2020mirror, yang2025humanai, liu2025theraissist, he2026mirror} \\
& Pure Quantitative & 11 (24.4\%) & \cite{perogamvros2025cathartic, ahmed2025design, premalatha2025personalized, ni2023focusing, laban2026robot, phairoh2025craftjai, juchli2026digital, mora2026refugio, kim2025gpt4, zhang2024autojournaling, jain2024role} \\
& Pure Qualitative & 3 (6.7\%) & \cite{hwang2026augmentiary, wang2025designing, nunez2025searching} \\
\midrule
\multirow{4}{*}{\makecell[l]{\textbf{Study}\\\textbf{Design}}} 
& Within-Subjects  & 35 (77.8\%) & \cite{nepal2024mindscape, zhang2025exploring, zisquit2025ai, held2025ai, perogamvros2025cathartic, dechant2025future, zhang2026gpt, ahmed2025design, honda2025communication, premalatha2025personalized, ni2023focusing, laban2026robot, phairoh2025craftjai, zhao2025research, ps2025neurosphere, vm2026serenitymind, zhu2026memorylib, yu2026mindfeed, hwang2026augmentiary, li2025healbubble, ikegami2026thank, wang2025designing, srivastava2026autismcarebot, burns2025remigraph, nunez2025searching, liu2026ragesense, zhong2026multimodal, park2025reimagining, juchli2026digital, mora2026refugio, rajcic2020mirror, yang2025humanai, zhang2024autojournaling, liu2025theraissist, he2026mirror} \\
& Between-Subjects & 7 (15.6\%) & \cite{kumar2025large, qin2024aiheal, liu2024moodmonster, park2025how, tsuyuguchi2026futureletter, nguyen2026beyond, kim2026using} \\
& Retrospective Observational Survey & 2 (4.4\%) & \cite{maples2026how, jain2024role} \\
& Expert \& Client Panel Evaluation & 1 (2.2\%) & \cite{kim2025gpt4} \\
\midrule
\multirow{2}{*}{\makecell[l]{\textbf{Duration}}} 
& Short-Term ($< 1 week$) & 17 (37.8\%) & \cite{zisquit2025ai, perogamvros2025cathartic, ahmed2025design, premalatha2025personalized, ni2023focusing, laban2026robot, zhao2025research, yu2026mindfeed, park2025how, li2025healbubble, srivastava2026autismcarebot, tsuyuguchi2026futureletter, burns2025remigraph, zhong2026multimodal, mora2026refugio, kim2026using, kim2025gpt4} \\
& Longitudinal ($\geq 1 week$) & 28 (62.2\%) & \cite{nepal2024mindscape, kumar2025large, zhang2025exploring, held2025ai, dechant2025future, zhang2026gpt, honda2025communication, phairoh2025craftjai, qin2024aiheal, liu2024moodmonster, ps2025neurosphere, vm2026serenitymind, zhu2026memorylib, hwang2026augmentiary, ikegami2026thank, wang2025designing, nunez2025searching, nguyen2026beyond, liu2026ragesense, park2025reimagining, juchli2026digital, maples2026how, rajcic2020mirror, yang2025humanai, zhang2024autojournaling, jain2024role, liu2025theraissist, he2026mirror} \\
\bottomrule
\end{tabular}%
}
\caption{Method paradigm, study design and duration of evaluation across reviewed articles.}
\vspace{-20pt}
\label{tab:evaluation_designs_taxonomy}
\end{table}

\subsubsection{Evaluation Methods and Study Design}
We analyzed the evaluation strategies of the reviewed corpus across three key dimensions: method paradigms, study designs, and durations, as shown in Tab.~\ref{tab:evaluation_designs_taxonomy}.

The evaluation methodologies were predominantly mixed-methods ($68.9\%$, $n = 31$). These studies coupled quantitative instruments (\eg, standardized psychological scales and usability surveys) with qualitative methods (\eg, semi-structured post-study interviews and open-ended reflections) to triangulate users' experiences. For instance, MindScape~\cite{nepal2024mindscape} coupled longitudinal weekly Ecological Momentary Assessment (EMA) surveys and usability scales with post-study qualitative interviews, \citet{honda2025communication} integrated automated physiological/facial emotion tracking with semi-structured interviews to evaluate robotic reappraisal support.
In contrast, $24.4\%$ of the papers ($n = 11$) conducted pure quantitative evaluations, utilizing methods such as longitudinal psychometric scale pre/post-tests~\cite{juchli2026digital}, algorithmic scoring consistency benchmarked against human consensus~\cite{perogamvros2025cathartic}, and automated ground-truth accuracy metrics derived from continuous screen sensing~\cite{zhang2024autojournaling}.
Pure qualitative investigations represented the smallest proportion ($6.7\%$, $n = 3$), capturing insights through in-depth multi-week field interviews in Augmentiary~\cite{hwang2026augmentiary}, first-person autobiographical design probes with reflective diary field notes~\cite{wang2025designing}, and collaborative autoethnographic reflexive thematic analysis~\cite{nunez2025searching}.

In terms of study design, the majority of the reviewed literature adopted within-subjects designs ($77.8\%$, $n = 35$), evaluating within-individual differences before and after the use of the proposed tools or comparing baseline versus intervention conditions. For instance, MindFeed~\cite{yu2026mindfeed} asked each participant to complete two consecutive daily journaling sessions, alternating between a traditional journaling baseline and the proposed AI-assisted system to evaluate differences in reflective depth.
Meanwhile, $15.6\%$ of the studies ($n = 7$) conducted between-subjects experiments. For example, FutureLetter~\cite{tsuyuguchi2026futureletter} randomly assigned participants to three parallel groups: a control group with pre-defined encouragement, an intervention group receiving LLM text suggestions, and another receiving reflective question prompts to contrast support strategies during writing inactivity. Likewise, \citet{nguyen2026beyond} conducted a 24-day field experiment comparing a treatment group that used an LLM chatbot for granular affect labeling with a control group that used standard self-monitoring to assess improvements in emotion regulation.
The remaining three studies used other paradigms: \citet{maples2026how} and \citet{jain2024role} deployed retrospective observational surveys ($4.4\%$, $n = 2$) to investigate long-term naturalistic user behaviors, while \citet{kim2025gpt4} employed an expert and client panel evaluation ($2.2\%$, $n = 1$) to benchmark LLM-generated clinical formulation reports.

Regarding evaluation duration, more than sixty percent of the corpus ($62.2\%$, $n = 28$) conducted longitudinal evaluations spanning one week or longer ($\ge 1\text{ week}$). These ranged from multi-week field trials, such as Socrates 2.0~\cite{held2025ai}, which deployed a 4-week naturalistic feasibility study with 61 participants, to multi-month investigations, such as MindScape~\cite{nepal2024mindscape}, which conducted an 8-week field deployment to track weekly EMA psychological trajectories across 20 college students.
Conversely, $37.8\%$ of the studies ($n = 17$) were short-term assessments ($< 1\text{ week}$). For example, \citet{zisquit2025ai} evaluated a VR self-talk system in a single-session lab trial, while AutoJournaling~\cite{zhang2024autojournaling} conducted a 5-day continuous screenshot deployment to benchmark automated journaling against daily ground-truth entries.

\begin{table}[t!]
\centering
\small
\setlength{\tabcolsep}{6pt}
\resizebox{\textwidth}{!}{
\begin{tabular}{>{\raggedright\arraybackslash}p{0.15\linewidth} >{\raggedright\arraybackslash}p{0.2\linewidth} >{\raggedright\arraybackslash}p{0.15\linewidth} >{\raggedright\arraybackslash}p{0.45\linewidth}}
\toprule
\textbf{Demographic Dimension} & \textbf{Category} & \textbf{\# of Studies (\%)} & \textbf{References} \\
\midrule
\multirow{6}{*}{\textbf{Sample Size ($N$)}} 
& $1 \le N \le 10$ & 8 (17.8\%) & \cite{ahmed2025design, laban2026robot, qin2024aiheal, wang2025designing, burns2025remigraph, nunez2025searching, kim2025gpt4, zhang2024autojournaling} \\
& $11 \le N \le 25$ & 19 (42.2\%) & \cite{nepal2024mindscape, zhang2025exploring, zisquit2025ai, honda2025communication, premalatha2025personalized, ni2023focusing, phairoh2025craftjai, liu2024moodmonster, vm2026serenitymind, zhu2026memorylib, yu2026mindfeed, hwang2026augmentiary, ikegami2026thank, srivastava2026autismcarebot, liu2026ragesense, zhong2026multimodal, park2025reimagining, rajcic2020mirror, liu2025theraissist} \\
& $26 \le N \le 50$ & 10 (22.2\%) & \cite{dechant2025future, zhao2025research, ps2025neurosphere, park2025how, li2025healbubble, nguyen2026beyond, mora2026refugio, yang2025humanai, jain2024role, he2026mirror} \\
& $51 \le N \le 100$ & 3 (6.7\%) & \cite{kumar2025large, held2025ai, tsuyuguchi2026futureletter} \\
& $N > 100$ & 4 (8.9\%) & \cite{perogamvros2025cathartic, juchli2026digital, maples2026how, kim2026using} \\
& Unreported & 1 (2.2\%) & \cite{zhang2026gpt} \\
\midrule
\multirow{5}{*}{\makecell[l]{\textbf{Age}}} 
& Young Adults (18--25 yrs) & 17 (37.8\%) & \cite{ahmed2025design, premalatha2025personalized, liu2024moodmonster, li2025healbubble, nguyen2026beyond, ikegami2026thank, park2025how, zisquit2025ai, honda2025communication, yu2026mindfeed, park2025reimagining, perogamvros2025cathartic, dechant2025future, zhu2026memorylib, tsuyuguchi2026futureletter, hwang2026augmentiary, zhong2026multimodal} \\
& Young to Middle-Aged Adults (18--40 yrs) & 16 (35.6\%) & \cite{held2025ai, zhang2025exploring, zhao2025research, qin2024aiheal, ps2025neurosphere, burns2025remigraph, liu2026ragesense, mora2026refugio, maples2026how, kim2026using, rajcic2020mirror, kim2025gpt4, yang2025humanai, jain2024role, he2026mirror, liu2025theraissist} \\
& Adolescents ($<18$ yrs) & 1 (2.2\%) & \cite{ni2023focusing} \\
& Older Adults ($>60$ yrs) & 1 (2.2\%) & \cite{laban2026robot} \\
& Unreported & 10 (22.2\%) & \cite{nepal2024mindscape, kumar2025large, zhang2026gpt, phairoh2025craftjai, vm2026serenitymind, wang2025designing, srivastava2026autismcarebot, nunez2025searching, juchli2026digital, zhang2024autojournaling} \\
\midrule
\multirow{9}{*}{\makecell[l]{\textbf{Mental Health}\\\textbf{State}}} 
& General Non-clinical & 32 (71.1\%) & \cite{nepal2024mindscape, kumar2025large, zhang2025exploring, zisquit2025ai, ahmed2025design, honda2025communication, premalatha2025personalized, ni2023focusing, laban2026robot, phairoh2025craftjai, qin2024aiheal, ps2025neurosphere, vm2026serenitymind, zhu2026memorylib, yu2026mindfeed, hwang2026augmentiary, ikegami2026thank, wang2025designing, tsuyuguchi2026futureletter, burns2025remigraph, nunez2025searching, nguyen2026beyond, zhong2026multimodal, park2025reimagining, mora2026refugio, maples2026how, kim2026using, rajcic2020mirror, yang2025humanai, zhang2024autojournaling, jain2024role, liu2025theraissist} \\
\cmidrule(lr){2-4}
& \makecell[l]{Non-clinical\\(Stressed/Trauma)} & 8 (17.8\%) & \makecell[l]{%
Childhood trauma: \cite{li2025healbubble}; \\
Academic / student stress: \cite{park2025how, dechant2025future}; \\
Workplace stress: \cite{he2026mirror}; \\
Gaming rage: \cite{liu2026ragesense}; \\
VR induced stress: \cite{zhao2025research}; \\
Mild psychological strain / distress: \cite{held2025ai}; \\
Autism-related support burden: \cite{srivastava2026autismcarebot}%
} \\
\cmidrule(lr){2-4}
& Sub-clinical / At-risk & 3 (6.7\%) & \makecell[l]{%
Nightmare disorder: \cite{perogamvros2025cathartic}; \\
Mental sub-health (stress/anxiety): \cite{liu2024moodmonster}; \\
Interpersonal distress (At-risk): \cite{kim2025gpt4}%
} \\
\cmidrule(lr){2-4}
& Unreported & 2 (4.4\%) & \cite{zhang2026gpt, juchli2026digital} \\
\bottomrule
\end{tabular}%
}
\caption{Summary of participant demographics across reviewed articles.}
\vspace{-25pt}
\label{tab:participant_demographics}
\end{table}

\subsubsection{Participant Demographics}
Table~\ref{tab:participant_demographics} summarizes the participant characteristics across the reviewed studies, encompassing sample size scales, age profiles, and mental health states. 

Across the corpus, $17.8\%$ of the studies ($n = 8$) conducted user evaluations with small cohorts ($1 \le N \le 10$), such as AutoJournaling~\cite{zhang2024autojournaling} ($N = 3$), deploying continuous mobile screenshot logging over five days to benchmark automated diary generation against manual ground-truth. 
The majority of evaluations ($42.2\%$, $n = 19$) were conducted with sample sizes between 11 and 25, such as Augmentiary~\cite{hwang2026augmentiary} ($N = 25$), which assessed reflective meaning-making through a 4-week in-the-wild deployment with interim and final interviews. 
An additional $22.2\%$ ($n = 10$) enrolled medium cohorts ($26 \le N \le 50$), exemplified by Refugio~\cite{mora2026refugio}, which recruited 30 participants to evaluate a VR emotional regulation safe-place. 
In comparison, larger-scale evaluations were the least frequent, with $6.7\%$ ($n = 3$) recruiting $51 \le N \le 100$ and $8.9\%$ ($n = 4$) engaging more than 100 participants. 
For example, Socrates 2.0~\cite{held2025ai} ($N = 61$) conducted a 4-week naturalistic feasibility trial using pre/post-intervention symptom scales, while ~\citet{juchli2026digital} ($N = 159$) tracked mental health scores across a 3-month field deployment. One study ($2.2\%$, $n = 1$) did not explicitly report its total sample size.

In terms of age distribution, the participants in the reviewed articles were generally young and working-age populations from universities.
Specifically, studies primarily recruited young adults aged 18–25 ($37.8\%$, $n = 17$), closely followed by young to middle-aged adults aged 18–40 ($35.6\%$, $n = 16$). Exemplifying the former, HealBubble~\cite{li2025healbubble} evaluated its emotion visualization app with 43 college students (aged 18–25) through demo interaction tasks. Representing the latter, RageSense~\cite{liu2026ragesense} deployed real-time gaming rage interventions in the wild with 53 mobile gamers aged 18–35 across three weeks.
Vulnerable age groups remained substantially underrepresented, with only one study focusing on adolescents under 18 ($2.2\%$) and one on older adults over 60 ($2.2\%$). As the sole study focusing on minors, ~\citet{ni2023focusing} evaluated a need-focused emotion regulation chatbot with 21 adolescents (Mean age = 14.7). 
Specifically addressing eldercare, ~\citet{laban2026robot} conducted a within-subjects comparative evaluation with two older adults (mean age = 85, range 80--90) to test an LLM-powered communication robot.
$22.2\%$ of the corpus ($n = 10$) did not report participant age demographics.

As for participants' mental health, the vast majority of studies ($71.1\%$, $n = 32$) evaluated systems with general non-clinical participants.
In contrast, $17.8\%$ of the corpus ($n = 8$) targeted non-clinical sub-populations undergoing specific emotional burdens or situational stressors. For example, HealBubble~\cite{li2025healbubble} engaged college students with histories of childhood trauma to assess trauma-informed emotion visualization, while ~\citet{park2025how} recruited university students experiencing severe academic stress to evaluate self-disclosing chatbot assessments. Other works in this category investigated gaming-induced rage~\cite{liu2026ragesense}, VR-induced emotional distress~\cite{zhao2025research}, preclinical psychological strain~\cite{held2025ai}, and support burdens among caregivers and educators of autistic individuals~\cite{srivastava2026autismcarebot}.
In addition, $6.7\%$ of the reviewed literature ($n = 3$) evaluated tools with sub-clinical or at-risk populations. Specifically, ~\citet{perogamvros2025cathartic} benchmarked LLM dream scoring against clinical patients diagnosed with nightmare disorder; MoodMonster~\cite{liu2024moodmonster} tracked daily mood trajectories among subclinical individuals experiencing chronic stress and anxiety; and ~\citet{kim2025gpt4} engaged at-risk clients experiencing interpersonal distress to evaluate AI-generated psychodynamic formulation reports. 
Finally, two studies ($4.4\%$, $n = 2$) did not report participants' mental health background.

\section{Discussion}
\subsection{Critical Reflection}
\label{sec: critical reflection}
\textbf{I. Information Loss and Error Accumulation in Text-Centric Processing.} 
A significant limitation across current LLM-based affective PI systems is their overwhelming reliance on text-centric processing. As synthesized in Tab.~\ref{tab:data_modalities} and Sec.~\ref{sec: system input modalities}, pure textual input constitutes the clear majority of reviewed implementations ($60.0\%$, $n = 27$). More crucially, most of the remaining 18 studies ($40.0\%$) that incorporated non-textual modalities (\eg, visual, speech, physiological, and behavioral data) did not employ LLMs to directly process raw sensory representations. Instead, they relied on multi-stage preprocessing pipelines that convert raw non-textual data into intermediate textual summaries, feeding only these textual proxies into downstream LLMs. \textit{This architectural paradigm inevitably leads to substantial loss of multimodal information and cascading error accumulation.}

Summarizing non-textual modalities into textual descriptions frequently results in the loss of original, fine-grained affective information~\cite{jordan2025speech, han2025benchmarking}. For instance, in speech-based interactions, standard ASR transcriptions strip away rich paralinguistic cues (\eg, pitch inflection, vocal tremors, and pauses) that often carry far more critical emotional information than semantic words alone~\cite{jordan2025speech}. Similarly, continuous physiological signals (\eg, HRV) contain abundant temporal dynamics~\cite{han2025benchmarking}; condensing them into static text prompts with discrete affective labels erases these continuous temporal trajectories.
As a result, LLMs are unable to align sudden physiological shifts with specific contextual triggers, which compromises the accuracy of reflective causal explanations~\cite{englhardt2024classification}.

Furthermore, such multi-stage pipelines are inherently susceptible to cumulatively cascading computational errors. For example, in MindScape~\cite{nepal2024mindscape}, geospatial clustering and activity recognition APIs aggregate raw behavioral data into textual summaries. However, imprecise clustering thresholds or recognition errors can cause inaccurate behavioral inferences to cascade directly into the prompt, misleading the LLM. Consequently, the LLM may generate inaccurate causal reasoning and subsequently issue unhelpful or incorrect recommendations, ultimately confusing users.
Similarly, AI-HEAL~\cite{qin2024aiheal} converts raw EEG biofeedback into discrete affective state estimates, where coarse-grained classification rules can mislabel a user's emotional state, causing the downstream LLM to base its entire cognitive reappraisal dialogue on an erroneous premise. At a broader scale, these technical bottlenecks propagate into broader application failures, preventing affective PI tools from delivering reliable, sustained psychological support in users' everyday lives.

To mitigate these limitations, future research must shift away from inadequate and fragile textual proxies toward native multimodal large language models (MLLMs) capable of jointly modeling textual and non-textual signals end-to-end, which we further discuss in Sec.~\ref{sec: future direction}, Part I.
\\

\noindent \textbf{II. Over-Reliance on Active Screen-Based Interactions.}
A prominent limitation across the reviewed articles is their pervasive reliance on active, screen-bound interactions, which places a significant interaction and cognitive burden on users~\cite{luger2016like, epstein2015lived}. As illustrated in Table~\ref{tab:tools_categories} and Sec.~\ref{sec: tool categories}, the vast majority of systems are instantiated as affective support \& intervention systems ($31.1\%$, $n = 14$), journaling support ($28.9\%$, $n = 13$), and chatbots ($20.0\%$, $n = 9$) deployed on smartphones or web interfaces. 
Across these tools, interaction paradigms remain overwhelmingly active and text-centric. To illustrate, users need to explicitly formulate verbose reflections in TrueBalance~\cite{zhang2025exploring}, type out introspective narratives in Augmentiary~\cite{hwang2026augmentiary}, or repeatedly engage with multi-turn Socratic questioning sequences in Socrates 2.0~\cite{held2025ai}.

This reliance on demanding active engagement creates a critical usability paradox: individuals undergoing acute emotional distress, burnout, or cognitive overload often experience severe executive function depletion~\cite{arnsten2009stress}, which greatly inhibits their capacity to perform complex text-based interactions or process lengthy AI feedback~\cite{epstein2015lived}. Forcing users to input complex text or read dense information on screens during vulnerable moments introduces secondary cognitive fatigue~\cite{kelley2017self}. Consequently, affective support tools risk becoming a digital burden that produces the opposite of the intended therapeutic effect.

In contrast, systems exploring physical embodiment and spatial immersion remain markedly under-explored, with embodied AI ($11.9\%$, $n = 5$) and VR environments ($7.1\%$, $n = 3$) collectively representing only $19.0\%$ ($n = 8$) of the corpus. Nonetheless, they demonstrate diverse possibilities to reduce active operational demands and screen reliance, such as implicitly collecting users' facial cues through smart mirrors (Mirror Ritual~\cite{rajcic2020mirror}), providing supportive physical co-presence guided by biofeedback signals (QTrobot~\cite{laban2026robot}), and dynamically synthesizing relaxing 3D safe places via voice commands (Refugio~\cite{mora2026refugio}).

Therefore, transitioning from demanding active, screen-bound affective support toward low-burden, implicit, and physically or spatially grounded interactions is critical to reducing cognitive load for vulnerable users. We provide future solutions in Sec.~\ref{sec: future direction}, Part II.
\\

\noindent \textbf{III. Skewed Participant Demographics and Campus Bubbles.}
A critical empirical gap across the reviewed literature lies in the pronounced demographic homogeneity and sampling skew in user evaluations. As summarized in Table~\ref{tab:participant_demographics}, evaluated cohorts were disproportionately concentrated among young adults aged $18$--$25$ ($37.8\%$, $n = 17$) and young-to-middle-aged cohorts aged $18$--$40$ ($35.6\%$, $n = 16$), which together account for $94.3\%$ of studies with reported age data ($73.4\%$ of the entire corpus), while emotionally vulnerable demographics across broader life stages—such as adolescents ($<18$ years, $2.2\%$, $n = 1$) and older adults ($>60$ years, $2.2\%$, $n = 1$)—remain substantially under-represented. Furthermore, participant recruitment was overwhelmingly confined to academic campus environments (\eg, university students, staff, and faculty).

This prevalent 'campus bubble' of participants in evaluations likely introduces severe bias into the evaluation results of existing systems. Specifically, university populations and young cohorts generally exhibit higher digital literacy, higher cognitive flexibility, and more frequent exposure to LLMs, enabling them to navigate product usage obstacles, prompt formulation barriers, and LLM hallucinations~\cite{cotton2024chatting, zamfirescu2023johnny}. 
Consequently, positive feasibility and effectiveness assessed within these digitally literate populations do not guarantee validity when translated to broader sociodemographic groups. Diverse occupational populations (\eg, blue-collar workers) and older adults often encounter steeper digital readiness gaps~\cite{van2014digital}, age-related perceptual declines~\cite{czaja2006factors}, and distinct psychological coping mechanisms~\cite{carstensen1999taking, lachman1998sense}, which cannot be adequately captured by evaluations on campus cohorts. Therefore, robust validation across heterogeneous demographics remains an open challenge, which we explain in Sec.~\ref{sec: future direction}, Part III.

\subsection{Future Directions}
\label{sec: future direction}
\textbf{I. From Textual Proxies to End-to-End Multimodal Modeling.}
To resolve the information loss and cascading errors inherent in multi-stage text-centric pipelines (as shown in Sec.~\ref{sec: critical reflection}, Part I), future affective PI systems must transition toward MLLMs that can natively process diverse modalities, \eg, visual, audio, and time-series data. For acoustic and visual streams, instead of relying on discrete ASR transcriptions (\eg, \citet{zisquit2025ai} and \citet{honda2025communication}) or isolated VLMs (\eg, MindFeed~\cite{yu2026mindfeed} and MemoryLib~\cite{zhu2026memorylib}), researchers can leverage emerging audio- and vision-enabled MLLMs, \eg, GPT-4o~\cite{gpt-4o}, Gemini 1.5 Pro~\cite{gemini1.5}. Processing continuous acoustic and visual cues alongside textual content end-to-end preserves rich affective nuances that are otherwise stripped away by intermediate textual proxies.

Time-series signals, \ie, physiological and behavioral data, represent another critical input for affective systems owing to their objectivity, passivity, and high temporal resolution. However, general-purpose MLLMs remain ill-equipped to process continuous sensor time-series, such as raw EEG, electrocardiogram (ECG), and inertial measurement unit (IMU) streams. Rather than collapsing these non-stationary signals into coarse-grained textual descriptions as seen in existing tools (\eg, QTrobot~\cite{laban2026robot} and AI-HEAL~\cite{qin2024aiheal}), future research can draw inspiration from emerging domain-specific sensor-language models. Recent advances demonstrate that continuous biosignals can be modeled natively at scale: for instance, SleepFM~\cite{thapa2024sleepfm} and \citet{abbaspourazad2023large} perform self-supervised pre-training on large-scale multimodal physiological datasets, while NeuroLM~\cite{jiang2025neurolm} and ECG-Chat~\cite{zhao2025ecgchat} employ cross-modal projectors to bridge raw biosignals directly with LLM latent spaces for conversational clinical reasoning. Similarly, sensor-language architectures such as SensorLM~\cite{zhang2026sensorlm} and SensorLLM~\cite{li2025sensorllm} map temporal sensor patches into unified semantic token spaces. Adapting these cross-modal alignment mechanisms to affective PI would enable LLMs to ingest continuous sensor dynamics as soft tokens alongside conversational dialogue, mitigating the information loss and cascading errors inherent in coarse textual proxies.

We also note that while upstream MLLMs demand continued technical innovation, affective interactive systems research simultaneously lacks sufficient engagement with cutting-edge, domain-specific sensor foundation models. Moving toward native end-to-end multimodal modeling requires not only paradigm shifts in affective interaction design and breakthroughs in upstream model architectures, but also active synergy and mutual reinforcement between both fields. For example, AI researchers can design end-to-end LLM architectures to directly process raw wearable data, while HCI designers can employ these architectures in affective PI applications and contribute subjective user reflections to align these architectures with real-world emotional contexts.
\\

\noindent \textbf {II. Towards Low-Burden Affective Interactions.}
As identified in our critical reflection (Sec.~\ref{sec: critical reflection}, Part II), the prevailing tools' reliance on demanding, active, screen-bound interactions creates a severe usability paradox by overloading (vulnerable) users during acute emotional distress.
To address this limitation, it is necessary to reduce users' workload and cognitive load when using the systems.
However, a certain degree of active user operation is indispensable in affective support systems, as emotional understanding requires users to actively externalize subjective experiences and give feedback to AI-generated analyses~\cite{nunez2025searching}. 
Therefore, blindly pursuing fully passive interaction is not feasible.
Instead, future research needs to find a balance between achieving low-burden interaction and ensuring the effectiveness and usability of tools.

To build this low-burden yet effective paradigm, researchers can draw upon established interaction design principles. We suggest three feasible paths: (1) \textit{micro-interaction}~\cite{epstein2015lived}, (2) \textit{peripheral interaction}~\cite{bakker2015peripheral}, and (3) \textit{adaptive interaction}~\cite{gajos2007automatically}.
First, future research can adopt micro-interaction, which involves lightweight and single-action engagements allowing users to complete a targeted action in just a few seconds~\cite{epstein2015lived}. For example, compared to multi-step textual workflows, lightweight one-tap mood check-ins or single-action visual anchors substantially lower manual input workload~\cite{intille2016muema}.
Second, peripheral interaction~\cite{bakker2015peripheral}, which enables users to interact with systems at the periphery of their attention without interrupting primary activities, also offers a promising approach to lightweight affective support. Specifically, moving beyond rigid screens toward tangible objects and environments-such as soft lighting transitions~\cite{snyder2015moodlight}, or graspable physical artifacts~\cite{zhou2024squeezable}-enables emotional offloading without requiring focal visual attention.
Third, researchers could consider systems that adaptively adjust to users' fatigue levels and emotional states. To illustrate, the widely used conversational agents in the reviewed papers can be designed to dynamically regulate their dialogue pacing and verbosity: delivering concise text during periods of acute depletion and gradually expand into open-ended, reflective dialogues only as users' cognitive capacity recovers.
\\

\noindent \textbf {III. Evaluating Across Heterogeneous and Vulnerable Demographics.}
As shown in our critical reflection (Sec.~\ref{sec: critical reflection}, Part III), the reliance on campus-based sampling introduces severe bias and leads to a failure in establishing ecological validity across real-world populations.
To overcome this limitation, future research must proactively expand participant recruitment beyond university bubbles to include underrepresented and emotionally vulnerable populations across diverse life stages. This also requires researchers to reconsider interface complexities, prompt formulation barriers, and system misunderstandings in the design of systems for less tech-literate or non-tech-literate populations.

Furthermore, empirical evaluations must account for heterogeneous socioeconomic backgrounds~\cite{naslund2017digital}, occupational strains~\cite{mark2008cost}, and differing psychological coping mechanisms~\cite{carver1989assessing}. Rather than relying on short-term laboratory trials with healthy participants, future investigations should conduct longitudinal in-the-wild studies~\cite{klasnja2011evaluate} involving individuals exposed to diverse real-world stressors, including blue-collar workers~\cite{jebelli2018feasibility} and subclinical populations~\cite{bakker2016mental}. Such inclusive evaluations will enable researchers to identify demographic-specific failure modes—such as clinical misinterpretations~\cite{yang2024mentallama}, cognitive friction from unadapted interaction pacing~\cite{gnewuch2018faster}, or cultural misalignments in LLM advice—thereby~\cite{aleem2024towards} thereby ensuring that affective PI tools deliver genuinely safe, robust, and equitable psychological support.

\subsection{Limitations}
Our scoping review exhibits several methodological limitations that warrant consideration. 
First, our search protocol was bounded by peer-reviewed English literature published between January 2020 and June 2026 across four established digital libraries. Consequently, commercial, closed-source consumer applications, emerging preprints, and non-English systems were excluded, which may omit certain industrial advances in generative affective tools. 
Second, given the rapid release cycles of LLMs, our analysis captures a specific evolutionary window; the model capabilities and architectural paradigms documented here will naturally continue to shift as native multimodal models mature. 
Finally, our synthesis relied primarily on authors' self-reported system implementations and empirical findings. Due to the scarcity of open-source artifacts, public benchmarking datasets, and standardized validation protocols across the reviewed corpus, we could not independently audit technical claims or benchmark quantitative performances across systems.

\section{Conclusion}
In this work, we presented a comprehensive scoping review of 45 research articles to systematically chart the emerging landscape of LLM-based affective PI tools. By mapping macro temporal and model adoption trajectories, establishing a multi-layered system taxonomies, and dissecting research methodologies across the system lifecycle, our results indicate that current LLM-based affective PI tools still encounter design and methodological challenges in pipeline construction, interaction modality, and participant recruitment. To address these issues, we provide actionable recommendations for future research to develop more robust, low-burden, and inclusive systems within the field.

\bibliographystyle{ACM-Reference-Format}
\bibliography{ref}

\appendix

\end{document}